\documentclass[referee]{aa} 

\usepackage{CJK}
\usepackage{graphicx}
\graphicspath{{figures/}}       
\usepackage{txfonts}
\usepackage{footnote}
\usepackage{hyperref}
\usepackage{natbib}

\usepackage{color}

\begin{document}

   \title{Ongoing star formation in the protocluster IRAS 22134+5834}
        \titlerunning{Protocluster IRAS 22134+5834}
         \authorrunning {Yuan Wang et al.}
   \author{Yuan Wang\inst{1}\inst{,2}, Marc Audard\inst{1}, Francesco Fontani\inst{3}, \'Alvaro S\'anchez-Monge\inst{4}, 
   Gemma Busquet\inst{5}\inst{,6}, Aina Palau\inst{7},  Henrik Beuther\inst{8}, Jonathan C. Tan\inst{9}, Robert Estalella\inst{10},
   Andrea Isella\inst{11}, Frederic Gueth\inst{12}, and Izaskun Jim\'enez-Serra\inst{13}
   }
 
   \institute{Department of Astronomy, University of Geneva,
              Chemin d'Ecogia 16, CH-1290 Versoix, Switzerland\\
              \email{yuan.wang@unige.ch}
         \and
             Purple Mountain Observatory, Chinese Academy of Sciences, 2 West Beijing Road, 210008 Nanjing, P.R. China\\
           \and
             INAF -- Osservatorio Astrofisico di Arcetri, L.go E. Fermi 5, 50125 Firenze, Italy\\
            \and
             I. Physikalisches Institut, Universit\"at zu K\"oln, Z\"ulpicher Str.\ 77, 50937, Cologne, Germany\\
            \and
            Insitut de Ci\`encies de l’Espai (CSIC-IEEC), Campus UAB, carrer de Can Magran S/N 08193, Cerdanyola del Vall\`es, Catalunya, Spain\\
            \and
            Instituto de Astrof\'isica de Andaluc\'ia, CSIC, Glorieta de la Astronom\'ia, s/n E-18008 Granada, Spain\\
            \and
           Instituto de Radioastronom\'ia y Astrof\'isica, Universidad Nacional Aut\'onoma de M\'exico, P.O. Box 3-72, 58090, Morelia, Michoac\'an, Mexico\\
            \and 
             Max-Planck-Institut f\"ur Astronomie, K\"onigstuhl 17, D-69117, Heidelberg, Germany\\
            \and
            Departments of Astronomy and Physics, University of Florida, Gainesville, FL 32611, USA\\
            \and
            Dept. d'Astronomia i Meteorologia, Institut de Ciencies del Cosmos, Univ. de Barcelona, IEEC-UB, Marti Franques 1, E08028 Barcelona, Spain\\
            \and
            Department of Physics \& Astronomy - MS 108, Rice University, 6100 Main Street, Houston, TX, 77005, USA\\
            \and
            Institut de Radioastronomie Millim\'etrique (IRAM), 300 rue de la Piscine, 38406 Saint Martin d’H\`eres, France\\
            \and
            University College London, Department of Physics and Astronomy, 132 Hampstead Road, London NW1 2PS, UK\\
            }{
             }

  \date{Received May 2015; accepted Oct. 2015}

 
  \abstract
   {}
   {Massive stars form in clusters, and their influence on nearby starless cores is still poorly understood. The protocluster associated with IRAS 22134+5834 represents an excellent laboratory for studying the influence of massive YSOs on nearby starless cores and the possible implications in the clustered star formation process.}
   {IRAS 22134+5834 was observed in the cm range with (E)VLA, 3~mm with CARMA, 2~mm with PdBI, and 1.3~mm with SMA, to study both the continuum emission and the molecular lines that trace different physical conditions of the gas.}
   {The multiwavelength centimeter continuum observations revealed two radio sources within the cluster, VLA1 and VLA2. VLA1 is considered to be an optically thin UCH{\sc ii} region with a size of 0.01~pc that sits at the edge of the near-infrared (NIR) cluster. The flux of ionizing photons of the VLA1 corresponds to a B1 ZAMS star. VLA2 is associated with an infrared point source and has a negative spectral index. We resolved six millimeter continuum cores at 2~mm, MM2 is associated with the UCH{\sc ii} region VLA1, and other dense cores are distributed around the UCH{\sc ii} region. Two high-mass starless clumps (HMSC), HMSC-E (east) and HMSC-W (west), are detected around the NIR cluster with N$_2$H$^+$(1--0) and NH$_3$ emission, and they show different physical and chemical properties. Two N$_2$D$^+$ cores are detected on an NH$_3$ filament close to the UCH{\sc ii} region with a projected separation of $\sim$8000~AU at the assumed distance of 2.6~kpc. The kinematic properties of the molecular line emission confirm that the UCH{\sc ii} region is expanding and that the molecular cloud around the near infrared (NIR) cluster is also expanding.}
   {Our multiwavelength study has revealed different generations of star formation in IRAS 22134+5834. The formed intermediate-to-massive stars show a strong impact on nearby starless clumps. We propose that the starless clumps and HMPOs formed at the edge of the cluster while the stellar
wind from the UCH{\sc ii} region and the NIR cluster drives the large scale
bubble. }


   \maketitle
%

\section{Introduction}
Even though most stars in our Galaxy form in rich clusters \citep[e.g.,][]{lada2003}, the initial conditions of clustered star formation are still poorly understood. Studies of low-mass prestellar cores in low-mass star-forming regions suggest that cluster environments have a relatively weak influence on the properties of the cores (i.e., temperature, mass, velocity dispersion, and chemical abundance, e.g., \citealt{andre2007,friesen2009, foster2009}). However, these conclusions probably do not apply to high-mass star-forming regions. Previous high-angular resolution observations have revealed that high-mass sources with different evolutionary stages, from prestellar cores to ultra-compact H{\sc ii} (UCH{\sc ii}) regions, coexist in the same star-forming complex close to each other, and the age spread within the cluster members could be as large as $\sim$2 to 3~Myr (e.g., \citealt{bik2012, wang2011, palau2010}). The strong ultraviolet (UV) radiation, massive outflows from the newly formed high-mass stars, or expanding H{\sc ii} regions can have a strong impact on the environment around high-mass prestellar cores detected interferometrically in dense gas tracers (e.g., N$_2$H$^+$, NH$_3$, \citealt{palau2010, sanchez-monge2013}). 

Previous studies show that these energetic phenomena could have major effects on prestellar dense cores. Millimeter interferometer observations performed by \citet{palau2007b} detected several prestellar candidates compressed by the expanding cavity driven by a massive young stellar objects (YSO) in IRAS~20343+4129, and the line widths are clearly non-thermal, unlike what is observed in clustered low-mass prestellar cores. In the protocluster associated with IRAS~05345+3157, \citet{fontani2009} find that the kinematics of two prestellar core candidates are influenced by the passage of a massive outflow; \citet{palau2007a} find that in IRAS~20293+3952, the UV radiation and outflows can also affect the chemistry of starless cores; but the deuteration of species like N$_2$H$^+$ and NH$_3$ seems to remain as high as in prestellar cores associated with low-mass star-forming regions \citep{fontani2008, fontani2009, busquet2010, gerner2015}. However, the aforementioned regions have L$_{\rm bol}\lesssim$5,~000~$L_\odot$, and the effects of a higher mass protostar on its surroundings, both on small ($<$5000~au) and large scales ($>$10~000~au), remain poorly constrained from an observational point of view. On the other hand, a near-infrared (NIR) photometry and spectroscopy toward the high-mass star-forming region RCW~34 \citep{bik2010} suggests that the low- and intermediate-mass stars formed first and produced the ``bubble'', while the O star formed later at the edge and induced the formation of the next-generation stars in the molecular cloud.

Situated at a distance of 2.6~kpc, IRAS 22134+5834 (hereafter I22134) has a luminosity of 1.2$\times10^4$~$L_\odot$ and is associated with a centimeter compact source of a few mJy \citep{sridharan2002}. At millimeter wavelengths, this region is dominated by an extended continuum source that peaks on the IRAS source \citep{chini2001, beuther2002b}. A massive molecular outflow is also detected with single-dish telescopes in CO \citep{dobashi1994, dobashi2001, beuther2002b} and HCO$^+$(1--0) \citep{lopez-sepulcre2010}, which indicates that one or more high-mass stars are forming. The NIR $K-$band images reveal a ring-shaped embedded cluster made of early- to late-B type young stars with a central dark region \citep{kumar2003}. Furthermore, multiple starless cores are found around the centimeter source and the NIR cluster \citep{busquet_phd, sanchez-monge_phd, sanchez-monge2013}. Therefore, the protocluster associated with I22134 represents an excellent laboratory to study the influence of massive YSOs on nearby starless dense cores.

In this paper we investigate the properties of the dense cores in I22134 and the relations between with the ambient molecular clouds and the more evolved cluster members. In Sect.~\ref{sec_obs} we describe the observations and data reduction. In Sect.~\ref{sec_results} we present the millimeter and centimeter continuum results and the properties of the starless clumps from molecular line observations. We discuss properties of the sources and interactions between the YSOs and the molecular cloud in Sect.~\ref{sec_dis} and summarize the main conclusions in Sect.~\ref{sec_sum}


\section{Observations and data reduction}
\label{sec_obs}

We observed the star-forming region I22134 with the (Expanded) Very Large Array (VLA\footnote{The Very Large Array (VLA) is operated by the National Radio Astronomy Observatory (NRAO), a facility of the National Science Foundation operated under cooperative agreement by Associated Universities, Inc.}), the Combined Array for Research in Millimeter-wave Astronomy (CARMA\footnote{Supports for CARMA construction were derived from the Gordon and Betty Moore Foundation, the Kenneth T. and Eileen L. Norris Foundation, the Associates of the California Institute of Technology, the states of California, Illinois, and Maryland, and the National Science Foundation. Ongoing CARMA development and operations are supported by the National Science Foundation under a cooperative agreement and by the CARMA partner universities.}), the Plateau de Bure Interferometer (PdBI), and the Submillimeter Array (SMA\footnote{The Submillimeter Array is a joint project between the Smithsonian Astrophysical Observatory and the Academia Sinica Institute of Astronomy and Astrophysics and is funded by the Smithsonian Institution and the Academia Sinica.}). Tables~\ref{tab_obs} and \ref{tab_mmobs} summarize the observations.

\subsection{(E)VLA}
The I22134 star-forming region was observed with the VLA at 6.0, 3.6, 1.3, and 0.7~cm wavelengths in the A, B, and C configurations of the array in 2007 and 2009 with 10-21 EVLA antennas in the array. The data reduction followed the VLA standard guidelines for calibration of high-frequency data, using the NRAO package AIPS. We complemented our continuum observations with 1.4~GHz continuum archival data from the NRAO VLA Sky Survey (NVSS) \citep{condon1998}. We list the main parameters of the observations in Table~\ref{tab_obs}. 

The EVLA was also used on April 23 and May 07, 2010 to observe the continuum emission at 6.0 and 3.6~cm in its D configuration with a bandwidth of 2$\times$256~MHz (up to 5 times better than the standard VLA observations). The data reduction followed the EVLA standard guidelines for continuum emission, using the Common Astronomy Software Applications package (CASA). Images were done with the CLEAN procedure in CASA with different robust parameters (robust = --5 for the EVLA 6.0 and 3.6~cm data, and =0 for the rest of the images), the resulting rms noise levels and synthesized beams are shown in Table~\ref{tab_obs}.

The NH$_3$ $(J,K)=(1,1)$ and $(2,2)$ inversion lines (project AK558) were obtained during May 1, 2003 in its D configuration with the VLA. The phase center was R.A. 22$^{\rm h}$15$^{\rm m}08^{\rm s}.099$ Dec. +58\degr49\arcmin10\farcs00 (J2000). Amplitude and phase calibration were achieved by regularly interleaved observations of the quasar $2148+611$. The bandpass was calibrated by observing the bright quasars 1229$+$020 (3C273) and 0319$+$415 (3C84). The absolute flux scale was set by observing the quasar 1331$+$305 (3C286), and we adopted a flux of 2.41~Jy at the frequencies of 23.69 and 23.72~GHz. The 4IF spectral line mode was used to observe the NH$_3(1,1)$ and $(2,2)$ lines simultaneously with two polarizations. The bandwidth was 3.1~MHz, with 128~channels that had a channel spacing of 24.2~kHz ($\sim0.3$~km~s$^{-1}$) centered at $v_{\rm LSR}=-18.3$~km~s$^{-1}$. The data was  reduced with the software package AIPS following the standard guidelines for calibrating high-frequency data. Imaging was performed using natural weighting, and the resulting rms and synthesized beams are listed in Table~\ref{tab_lines}. The same data were also presented in \citet{sanchez-monge2013}.

\begin{table*}
\caption{\label{tab_obs} (E)VLA continuum observational parameters .}
\centering
\begin{tabular}{lcccccccc}
\hline\hline
\noalign{\smallskip}
$\lambda$ & Project ID.&   Config. & Epoch   &  Bootstr. flux of & Flux                        & Beam            & P.A.& rms                      \\
(cm)           &                  &                &              & Gain cal. 2148+611\tablefootmark{a}   &  cal.&  ($\arcsec \times \arcsec$)&(\degr)&($\mu$Jy~beam$^{-1}$)\\
\hline
\noalign{\smallskip}
VLA     &                     &                  &                 &               &         &          &  &\\
\hline
\noalign{\smallskip}
20.0   &NVSS\tablefootmark{b}  &   D &1995 Mar. 12  & ... &        ... &   45.$\times$45 &0& 470 \\
6.0     & AS902       &   A                         & 2007 Jun. 05  &1.28$\pm$1\%& 3C286& 0.41$\times$0.36&--17 & 37       \\
3.6  & AS902        &   A                        &2007 Jun. 05 & 1.05$\pm$1\% &3C286 &0.24$\times$0.19&$+$25&17\\
1.3  & AB1274      &   B                         &2007 Oct. 23& 0.72$\pm$3\%&3C286 &0.30$\times$0.22&$+31$&64\\
0.7  &  AS981       &   C                        &2009 Jun. 28 &0.68$\pm$5\% &3C286 &0.63$\times$0.39&$+68$&180\\
\hline
\noalign{\smallskip}
EVLA     &                     &                  &                 &               &         &          &  &\\
\hline
\noalign{\smallskip}
6.0     & AS1038     &   D                       & 2010 Apr 23& 1.29$\pm$1\%& 3C48&14.3$\times$10.0& $+33$&26 \\
3.6     & AS1038     &   D                       &2010 May 07&1.02$\pm$1\%& 3C48&8.3$\times$5.9     &+44 & 27\\
\hline
\end{tabular}
\tablefoot{
\tablefoottext{a}{Bootstraped flux is in Jy.}\\
\tablefoottext{b}{The detailed description for NVSS data can be found in \citet{condon1998}.}
}
\end{table*}

\subsection{CARMA}
CARMA was used to observe the 3~mm continuum and the selected molecular lines toward I22134 in the D configuration with 14 antennas (five 10.4~m antennas and nine 6~m antennas) in two different runs carried out on June 23, 2008 and May 4, 2010. During the first run, we covered the N$_2$H$^+(1-0)$ transition. The phase center was R.A. 22$^{\rm h}$15$^{\rm m}09^{\rm s}23$ Dec. $+58\degr49\arcmin08\farcs9$ (J2000),  and the baseline ranged from 10 to 108~m. The full width at half maximum (FWHM) of the primary beam at 93~GHz was 132$\arcsec$ for the 6~m antennas and 77$\arcsec$ for the 10.4~m antennas. The system temperatures were around 200~K during the observations. All the hyperfine transitions of N$_2$H$^+$(1--0) were covered with two overlapping 8~MHz units and 63 channels of each unit, resulting in a velocity resolution of $\sim$0.42~km~s$^{-1}$. Amplitude and phase calibration were achieved by regularly interleaved observations of BL Lac. The bandpass was calibrated by observing the brightest quasar 3C454.3. The absolute flux scale was set by observing MWC349 with an estimated uncertainty around 20\%. The observations carried out on 2010 were obtained with the new CARMA correlator, which provides up to eight bands. The phase center was R.A. 22$^{\rm h}$15$^{\rm m}08^{\rm s}1$ Dec. $+58\degr49\arcmin10\farcs0$ (J2000), and the baseline ranged from 9 to 95~m. We used three 500 MHz bands to observe the continuum, and five bands were set up to observe the C$_2$H($N=$1--0), CH$_3$OH($2_{0, 2}-1_{0, 1})$A+ (together with CH$_3$OH($2_{-1, 2}-1_{-1, 1})$E), HCO$^+$(1--0), {\it ortho}-NH$_2$D$(1_{1,1}-1_{0,1})$, CCS($7_6-6_5$) and c-C$_3$H$_2(2_{(1,2}-1_{0,1})$ line emission. Amplitude and phase calibration were achieved by regularly interleaved observations of the nearby quasar 2038+513. The bandpass was calibrated by observing 3C454.3, and the flux calibration was set by observing Neptune. The estimated uncertainty of the absolute flux calibration is $\sim10\%-15\%$.

Data calibration and imaging (natural weighting) were conducted in MIRIAD \citep{sault1995}. The resulting synthesized beam and rms for the continuum and detected molecular lines are listed in Tables~\ref{tab_mmobs} and~\ref{tab_lines}, respectively. We did not detect any emission from HCO$^+$(1--0),  CCS($7_6-6_5$), and c-C$_3$H$_2(2_{1,2}-1_{0,1})$ at the 3$\sigma$ limit of 130~mJy~beam$^{-1}$ with a velocity resolution of  0.4~km~s$^{-1}$, so we do not discuss these data in this paper.

\subsection{PdBI}
Observations in the 2~mm continuum and the N$_2$D$^+$(2--1) line at 154.217~GHz were obtained on June 2 and November 28, 2010, with the PdBI in the D and C configurations, respectively. The phase center was R.A. 22$^{\rm h}$15$^{\rm m}$10$^{\rm s}$.00 Dec. +58\degr49\arcmin02\farcs6 (J2000). Baselines range from $\sim$19--178~m. Atmospheric phase corrections were applied to the data, and the averaged atmospheric precipitable water vapor was $\sim$5~mm. The bandpass calibration was carried out using 1749+096. Phase and amplitude of the complex gains were calibrated by observing the nearby quasars 2146+608 and 2037+511. The phase rms is 10\degr to 40\degr, and the amplitude rms is below 10$\%$. The adopted flux density for the flux calibrator, MWC349, was 1.45~Jy. Calibration and imaging (natural weighting for N$_2$D$^+$(2--1) and robust = --2 for the continuum) were performed using the CLIC and MAPPING software of the GILDAS\footnote{The GILDAS software is developed at the IRAM and the Observatoire de Grenoble and is available at http://www.iram.fr/IRAMFR/GILDAS.} package with the standard procedures. The resulting synthesized beam and rms for the continuum and molecular lines are listed in Tables~\ref{tab_mmobs} and \ref{tab_lines}, respectively.

\subsection{SMA}
The source I22134 was observed with the SMA on September 28, 2010 in the extended configuration with eight antennas (baselines range $\sim$21--180~m). The phase center of the observations is R.A. 22$^{\rm h}$15$^{\rm m}10^{\rm s}.00$ Dec. +58\degr49\arcmin02\farcs6 (J2000), and the local standard of rest velocity was assumed to be = $-18.3$~km~s$^{-1}$. The SMA has two sidebands separated by 10~GHz. The correlator was set to 2~GHz mode, and the receivers were tuned to 231.321~GHz in the upper sideband and a uniform velocity resolution of $\sim0.6$~km~s$^{-1}$. The system temperatures were around $\sim$100--200~K during the observations (zenith opacities $\tau_{\rm 225~GHz}\lesssim0.08$ as measured by the Caltech Submillimeter Observatory). Bandpass was derived from the quasar 3C84 observations. Phase and amplitude were calibrated with regularly interleaved observations of quasars BL~Lac and 3C418. The flux calibration was derived from BL~Lac observations, and the flux scale is estimated to be accurate within 20$\%$. We applied different robust parameters for the continuum and line data (robust = 1 for the continuum, and 2 for the line images), the resulting synthesized beam and rms for the continuum and molecular lines are listed in Tables~\ref{tab_mmobs} and \ref{tab_lines}, respectively.  Besides the SO$(6_5-5_4)$ and CO(2--1), the two isotopologues of CO ($^{13}$CO and C$^{18}$O) were also detected; however, the emission of these two transitions suffered severe missing flux problem, and we do not discuss these data in this paper. The flagging and calibration were done with the IDL superset MIR \citep{scoville1993}, which was originally developed for the Owens Valley Radio Observatory and adapted for the SMA\footnote{The MIR cookbook by Chunhua Qi can be found at  \url{http://cfa-www.harvard.edu/~cqi/mircook.html}.}. The imaging and data analysis were conducted in MIRIAD \citep{sault1995}.

\begin{table*}
\caption{\label{tab_mmobs} Millimeter continuum observational parameters .}
\centering
\scriptsize
\begin{tabular}{lccccccccc}
\hline\hline
\noalign{\smallskip}
Array & $\lambda$&Config. & Epoch   & Gain&Bandpass &Flux& Beam\tablefootmark{$\ast$}       & P.A.\tablefootmark{$\ast$}& rms \tablefootmark{$\ast$} \\
             &  (mm) &                    &               &     cal.&   cal.        &cal.&   ($\arcsec \times \arcsec$)&(\degr)&(mJy~beam$^{-1}$)\\
\hline
\noalign{\smallskip}
CARMA  &   3  & D         &2008 Jun. 12 &BL Lac  &  3C454.3&   MWC349& 7.15$\times$5.71&$+$23& 0.38 \\
CARMA  &  3 & D           &2010 May 04&  2038+513 &  3C454.3&Neptune& 7.15$\times$5.71&$+$23& 0.38\\
PdBI    &   2  & D            & 2010 Jun. 02&  2146+608, 2037+511 &  1749+096&MWC349& 2.67$\times$1.58&$+$89& 0.26\\
PdBI    &   2  &C             & 2010 Nov. 28&  2146+608, 2037+511 &  1749+096&MWC349& 2.67$\times$1.58&$+$89& 0.26\\
SMA    &  1.3&Extended & 2010 Sep. 28& BL Lac, 3C418 &  3C84&BL Lac& 1.26$\times$1.00&$-$87& 0.50\\
\hline
\end{tabular}
\tablefoot{
\tablefoottext{$\ast$}{The beam size and and rms listed here for CARMA and PdBI observations are derived by combining the data from both configurations.}
}
\end{table*}

\begin{table*}
\caption{\label{tab_lines} Molecular line observation parameters.}
\centering
\begin{tabular}{lccccrcc}
\hline\hline
Lines&  Frequency\tablefootmark{a}& Telescope &  Config. & Beam     & PA & Velocity resol.& rms\tablefootmark{b}\\
        &       (GHz)   &                  &   &($\arcsec \times \arcsec$)   &(\degr)&(km~s$^{-1}$)&(mJy~beam$^{-1}$)\\
\hline
\noalign{\smallskip}
NH$_3(1,1)$&23.694500&   VLA &          D   & $3.77\times3.10$&+89 & 0.3 & 1.2\\
NH$_3(2,2)$&{ 23.722630}&   VLA &               D   & $3.69\times3.00$&+84 & 0.3 & 1.2\\
N$_2$H$^+$(1--0)&{93.176254}&CARMA&D& $5.49\times4.09$  &--82&0.4&31\\
C$_2$H($N=$1--0)&{ 87.316898}&  CARMA   &D &$6.49\times5.23$&+22&0.4&50\\
CH$_3$OH($2_{-1, 2}-1_{-1, 1})$E&{ 96.739362}&CARMA&D&$6.03\times4.79$&+27&0.06&108\\
CH$_3$OH($2_{0, 2}-1_{0, 1})$A+&{ 96.741375}&CARMA&D&$6.03\times4.79$&+27&0.06&108\\
{\it ortho}-NH$_2$D$(1_{1,1}-1_{0,1})$    &{ 85.926278}&CARMA&D&$6.79\times5.40$&+27&0.07&91\\
N$_2$D$^+$(2--1) &{  154.217011} &PdBI&C+D &$2.96\times1.94$&+94&0.3&6.0\\
CO(2--1)                &{ 230.538000}  &SMA&  Extended&$1.41\times1.02$&--88&0.6& 36 \\
SO$(6_5-5_4)$&{ 219.949442} &SMA&  Extended&$1.49\times1.07$&--88&0.6& 34 \\
\hline
\end{tabular}
\tablefoot{
\tablefoottext{a}{The frequencies for the lines were obtained from the Cologne Database for Molecular Spectroscopy (CDMS, \citealt{muller2001, muller2005}), except for NH$_3$ lines whose frequencies were obtained from the JPL Molecular Spectroscopy database \citep{pickett1998}.}\\
\tablefoottext{b}{The rms listed here is per channel.}
}
\end{table*}

\section{Results}
\label{sec_results}
\subsection{Centimeter continuum emission}
\label{sec_cmcont}
\begin{figure*}
   \centering
   \includegraphics[width=1\linewidth]{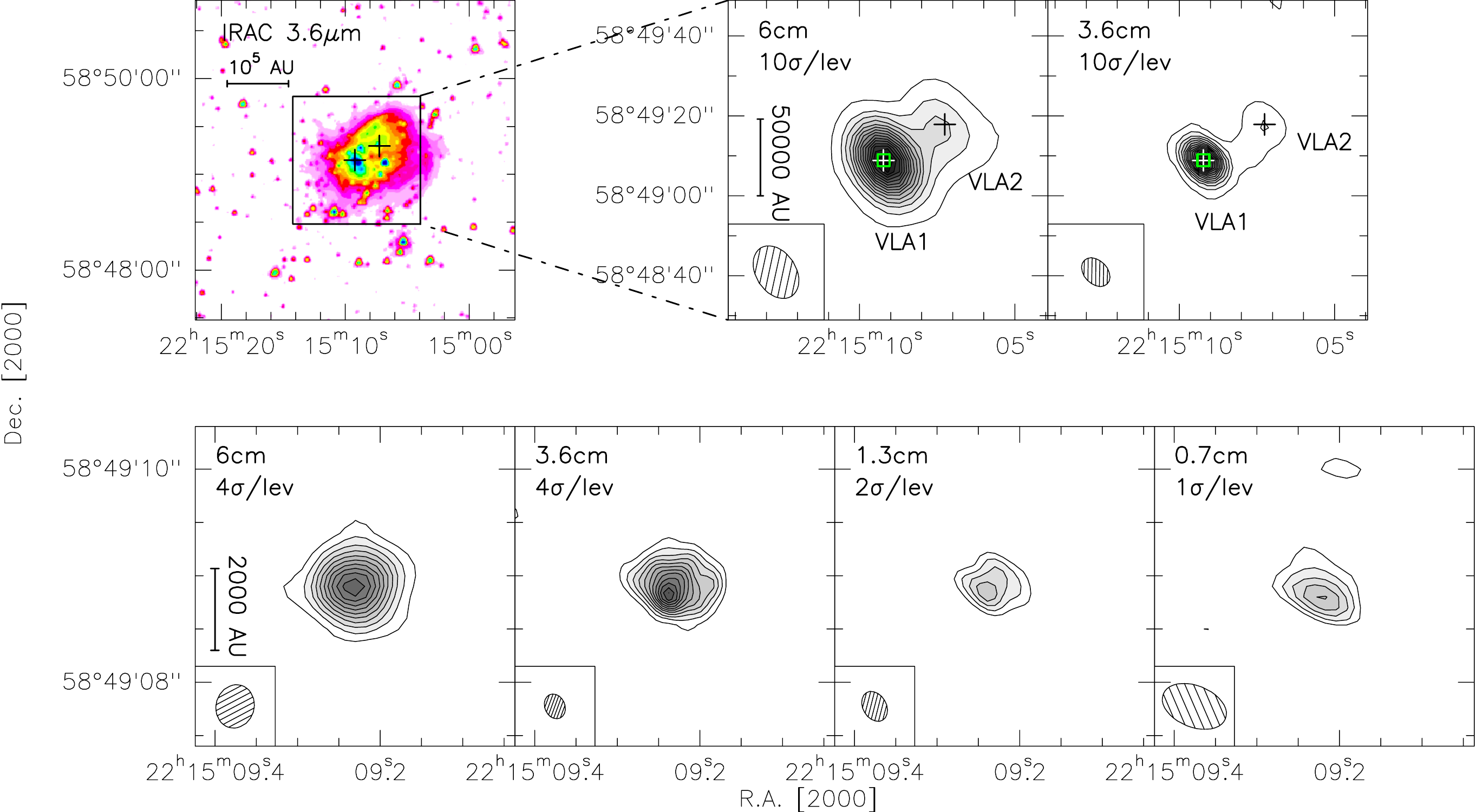}
   \caption{ {\it Spitzer}/IRAC and (E)VLA continuum emission images. {\it Top: Spitzer}/IRAC 3.6~$\mu$m, ELVA (D-config) 6.0~cm and 3.6~cm continuum images.  {\it Bottom:} The VLA 6.0~cm (A-config), 3.6~cm (A-config), 1.3~cm (B-config), and 0.7~cm (C-config) continuum images. The boxes in the EVLA continuum images in the top indicate the zoomed region in bottom panels. All the contour levels start at 4$\sigma$, and the steps are shown in each panel. The $\sigma$ values are listed in Table~\ref{tab_vla}. The crosses in the top panels mark the two radio sources we detected. The synthesized beams are shown in the bottom left corner of each panel. The {\it Spitzer}/IRAC post-bcd data processed with pipeline version S18.25.0 have been downloaded from the {\it Spitzer} archive to create the image.}
     \label{fig_cmcont}
\end{figure*}

VLA radio continuum emission is detected toward I22134 at all wavelengths (Fig. \ref{fig_cmcont}). At low angular resolution ($6\arcsec-14\arcsec$), the 6~cm and 3.6~cm emission is dominated by a strong and compact source VLA1, which is associated with one of the strongest IRAC sources in the field at 3.6~$\mu$m and is saturated at longer wavelengths in the IRAC band. A secondary radio source VLA2 is located $\sim15\arcsec$ to the northwest of VLA1 and is also associated with a B star \citep{kumar2003}. Only VLA1 is detected in the high angular resolution observations ($\lesssim 1\arcsec$, bottom panels in Fig. \ref{fig_cmcont}), and at 3.6~cm and 1.3~cm it is resolved into a cometary morphology with the head of the cometary arc pointing toward the southeast. The peak intensity and flux density measurements of VLA1 and VLA2 are listed in Table \ref{tab_vla}. The flux density of VLA1 is almost constant at all the VLA observations, while VLA2 shows a negative spectral index. 

\begin{table*}
\caption{Centimeter continuum properties of VLA1 { and VLA2}.}
\label{tab_vla} 
\centering
\begin{tabular}{lcccc}
\hline\hline
\noalign{\smallskip}
$\lambda$ & Beam                                & P.A.&$I_\nu$&$S_\nu$ \\
(cm)           & (\arcsec$\times$\arcsec)&(\degr)&(mJy~beam$^{-1}$)&(mJy)\\
\hline
\noalign{\smallskip}
VLA1  & 22:15:09.23 & +58:49:08.9  &{ (J2000)}& \\
\hline
\noalign{\smallskip}
20.0   & 45$\times$45                   &0                               &1.3$\pm$0.47&2.9$\pm$0.70\\
6.0     & 14.3$\times$10.0                      &$+33$                  &4.2$\pm$0.03&4.6$\pm$0.08\\
6.0     & 0.41$\times$0.36                      &--17                   &1.6$\pm$0.03&3.5$\pm$0.11\\
3.6     & 8.3$\times$5.9                        &$+44$                  &3.6$\pm$0.02&4.1$\pm$0.08\\
3.6  & 0.24$\times$0.19                 &$+$25                  &0.8$\pm$0.02&3.5$\pm$0.09\\
1.3  & 0.30$\times$0.22                 &$+31$                  &0.8$\pm$0.06&2.1$\pm$0.20\\
0.7  &  0.63$\times$0.39                        &$+68$                  &1.5$\pm$0.18&2.1$\pm$0.36\\
\hline
\noalign{\smallskip}
VLA2 &  22:15:07.26 & +58:49:17.9  &{ (J2000)}&\\
\hline
\noalign{\smallskip}
6.0     & 14.3$\times$10.0                      &$+33$                  &0.7$\pm$0.1&0.8$\pm$0.05 \\
3.6     & 8.3$\times$5.9                        &$+44 $                 &0.4$\pm$0.1&0.6$\pm$0.06\\
\hline
\end{tabular}
\tablefoot{
\tablefoottext{$\ast$}{Error in intensity is the rms noise level of the map. Error in flux density is estimated as $\sqrt{(\sigma\sqrt{\theta_{\rm source}/\theta_{\rm beam}})^2+(\sigma_{\rm flux-scale})^2}$, where $\theta_{\rm source}$ is the size of the source which we used to calculate the flux density (outlined by the lowest contour in Fig.~\ref{fig_cmcont}), $\theta_{\rm beam}$ is the size of the synthesized beam, $\sigma_{\rm flux-scale}$ is the error in the flux scale which takes into account the uncertainty of the calibration and is estimated as $S_\nu\times\%_{\rm uncertainty}$ ($\%_{\rm uncertainty}$ is listed in Table~\ref{tab_obs}).}
}
\end{table*}

\begin{figure}
   \centering
   \includegraphics[width=8cm]{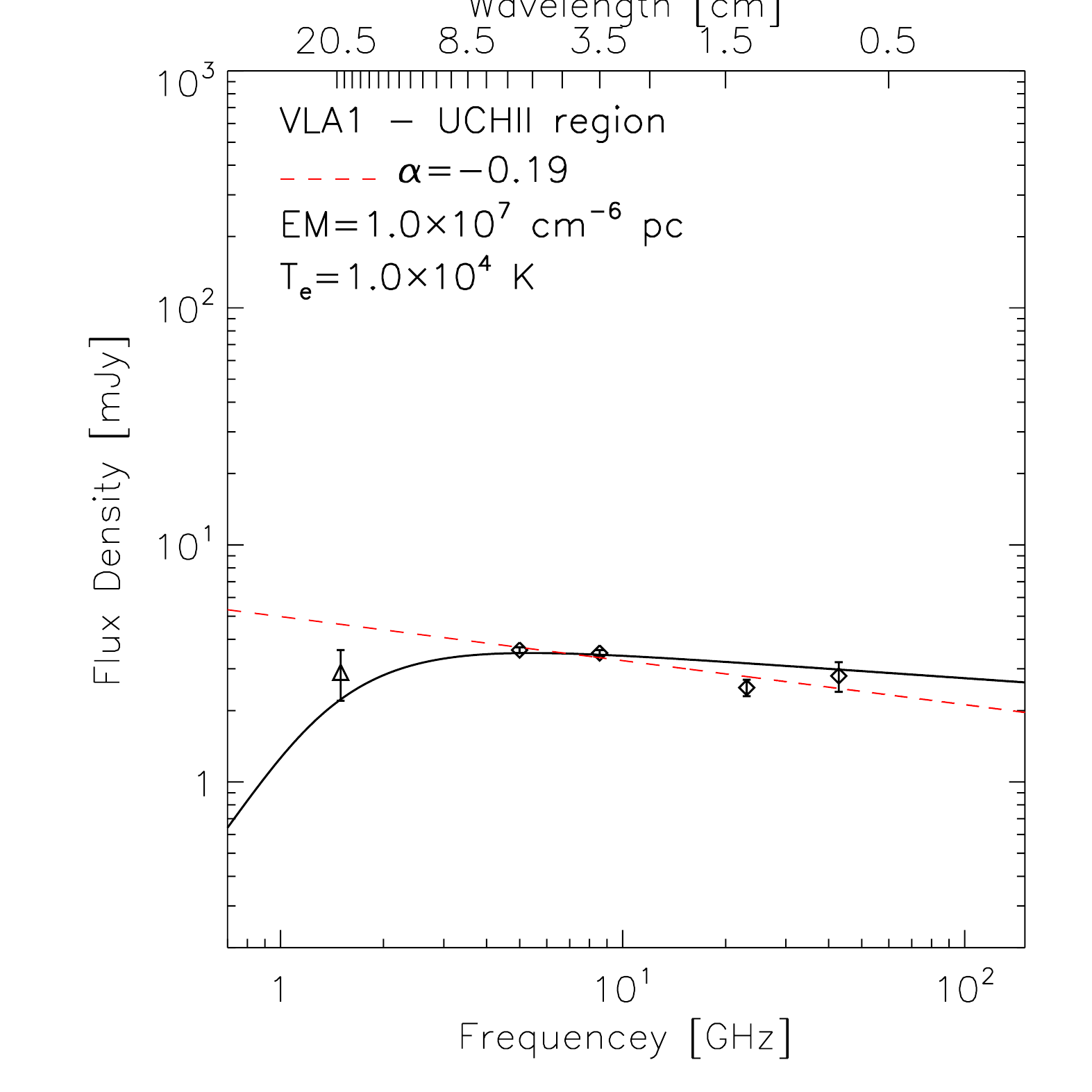}
   \caption{Spectral energy distribution for radio source VLA1. Diamonds correspond to the VLA high-angular resolution flux measurements listed in Table~\ref{tab_cm}. The triangle corresponds to the 20~cm flux measurement in Table~\ref{tab_vla}. The dashed line shows the linear fit ($S_\nu\propto\nu^\alpha$) to the high-angular resolution data (diamonds) with $\alpha=-0.19$. The solid line shows a homogeneous H{\sc ii} region fit to all the data points in the figure. }
     \label{fig_sed}
\end{figure}

To investigate the physical properties of VLA1, we reproduced the VLA images considering only the common $uv-$range (15--450~k$\lambda$) and convolved the resulting images to the same circular beam of 0\farcs7$\times$0\farcs7. The measured flux density and the deconvolved size of the continuum source VLA1 obtained for the images with the same uv range are listed in Table~\ref{tab_cm} (see also \citealt{sanchez-monge_phd}). Figure~\ref{fig_sed} shows the spectral energy distribution (SED) of the radio source VLA1. The flux of the source remains basically constant at centimeter wavelengths with a spectral index $\alpha=-0.19$ ($S_\nu\propto\nu^\alpha$) between 6 and 0.7~cm (Fig.~\ref{fig_sed}). Combining the 20~cm data, we can fit the SED with a homogeneous optically thin H{\sc ii} region with a size of 0.01~pc, an electron density ($n_{\rm e}$) of 5.5$\times10^4$~cm$^{-3}$, an emission measure (EM) of 1.0$\times10^7$~cm$^{-6}$~pc, an ionized gas mass of 9.1$\times10^{-5}$~$M_\odot$, and an ionizing photon flux of 2.5$\times10^{45}$~s$^{-1}$ (corresponding to a B1 ZAMS star, \citealt{panagia1973}). The size of VLA1 suggests that it could be a hyper-compact H{\sc ii} (HCH{\sc ii}) region; however, HCH{\sc ii} regions should have EM$\sim$1.0$\times10^{10}$~cm$^{-6}$~pc or $n_{\rm e}\sim5.5\times10^6$~cm$^{-3}$, which means the emission is typically optically thick at cm wavelengths \citep[spectral index $\alpha=2$,][]{kurtz2005, churchwell2002}, thus VLA1 is considered as a typical UCH{\sc ii} region. { The definitions of HCH{\sc ii} and UCH{\sc ii} used here differ from those of \citet{beuther2007} and \citet{tan2014}, where HCH{\sc ii}s are defined primarily based on size, i.e., as being $<$0.01pc in diameter, and similarly UCH{\sc ii}s as being $<$0.1pc in diameter.} It is worth noticing that at 20~cm, we did not exclude the emission of VLA2, so the fitting result is derived with a 20~cm flux that could be overestimated. As shown in Table~\ref{tab_vla}, at 3.6 and 6.0~cm the flux of VLA2 is only about 1/6 of VLA 1. If VLA2 shares the similar SED to VLA1 as we show in Fig.~\ref{fig_sed}, about one-seventh of the total emission at 20~cm might be due to VLA2. If VLA2 keeps the same power law spectral index between 3.6 and 6.0~cm to 20~cm, about one half of the total emission at 20~cm might be due to VLA2. When we removed the potential influence of VLA 2 to the 20~cm flux, the fit we obtained did not change (less than 1\%) and the shape of the SED is basically the same.

\begin{table*}
\caption{\label{tab_cm} Flux density for VLA1 with the same uv-range and same circular beam of 0\farcs7$\times$0\farcs7.}
\centering
\begin{tabular}{lcccc}
\hline\hline
\noalign{\smallskip}
$\lambda$ &$I_\nu$\tablefootmark{$\ast$}&    $S_\nu$\tablefootmark{$\ast$}&Deconv. size & P.A. \\
(cm)     &(mJy~beam$^{-1}$)&(mJy)   &  ($\arcsec \times \arcsec$)&(\degr)  \\
\hline
\noalign{\smallskip}
6.0             &2.7$\pm$0.04&                          3.6$\pm$0.1           &0.48$\times0.38\pm0.05$& 80$\pm$10        \\
3.6             &2.7$\pm$0.04&                          3.5$\pm$0.1            &0.44$\times0.34\pm0.05$& 80$\pm$10       \\
1.3             &2.0$\pm$0.07&                          2.5$\pm$0.2           &0.44$\times0.32\pm0.1$& 85$\pm$25 \\
0.7             &1.9$\pm$0.17&                          2.8$\pm$0.4            &0.62$\times0.37\pm0.3$& 85$\pm$30        \\
\hline
\end{tabular}
\tablefoot{
\tablefoottext{$\ast$}{The errors are estimated with the same method as described in the note of Table~\ref{tab_vla}.}
}

\end{table*}

\subsection{Millimeter continuum emission}
\label{sec_mmcont}
\begin{figure*}[ht]
   \centering
   \includegraphics[width=1\linewidth]{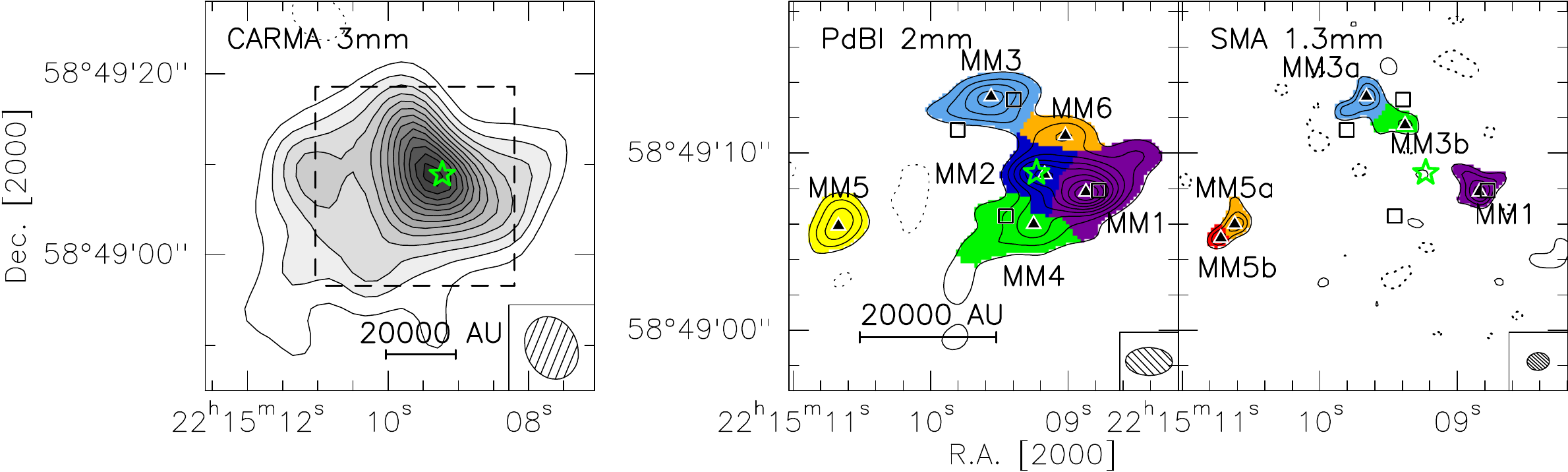}
   \caption{Millimeter continuum map obtained with the CARMA (left), PdBI (middle), and SMA (right). Contour levels start at 3$\sigma$ in steps of 2$\sigma$. The $\sigma$ of the contours in each panel, {\it from left to right}, is 0.38, 0.26, and 0.50~mJy~beam$^{-1}$, respectively. The dashed box in the left panel indicates the region plotted in the middle and right panels. The color scales in the PdBI and SMA panels show the core structures and boundaries derived from the ClumpFind procedure, the triangles and the numbers label the peaks as listed in Table~\ref{tab_cont}. The star marks the UCH{\sc ii} region VLA1. The squares indicate the cores detected by \citet{palau2013}. The dotted contours are the negative features due to the missing flux with the same contour levels as the positive ones in each panel. The synthesized beam is shown in the bottom right corner of each panel. }
     \label{fig_mmcont}
\end{figure*}

\begin{figure}[htbp]
   \centering
   \includegraphics[width=1\linewidth]{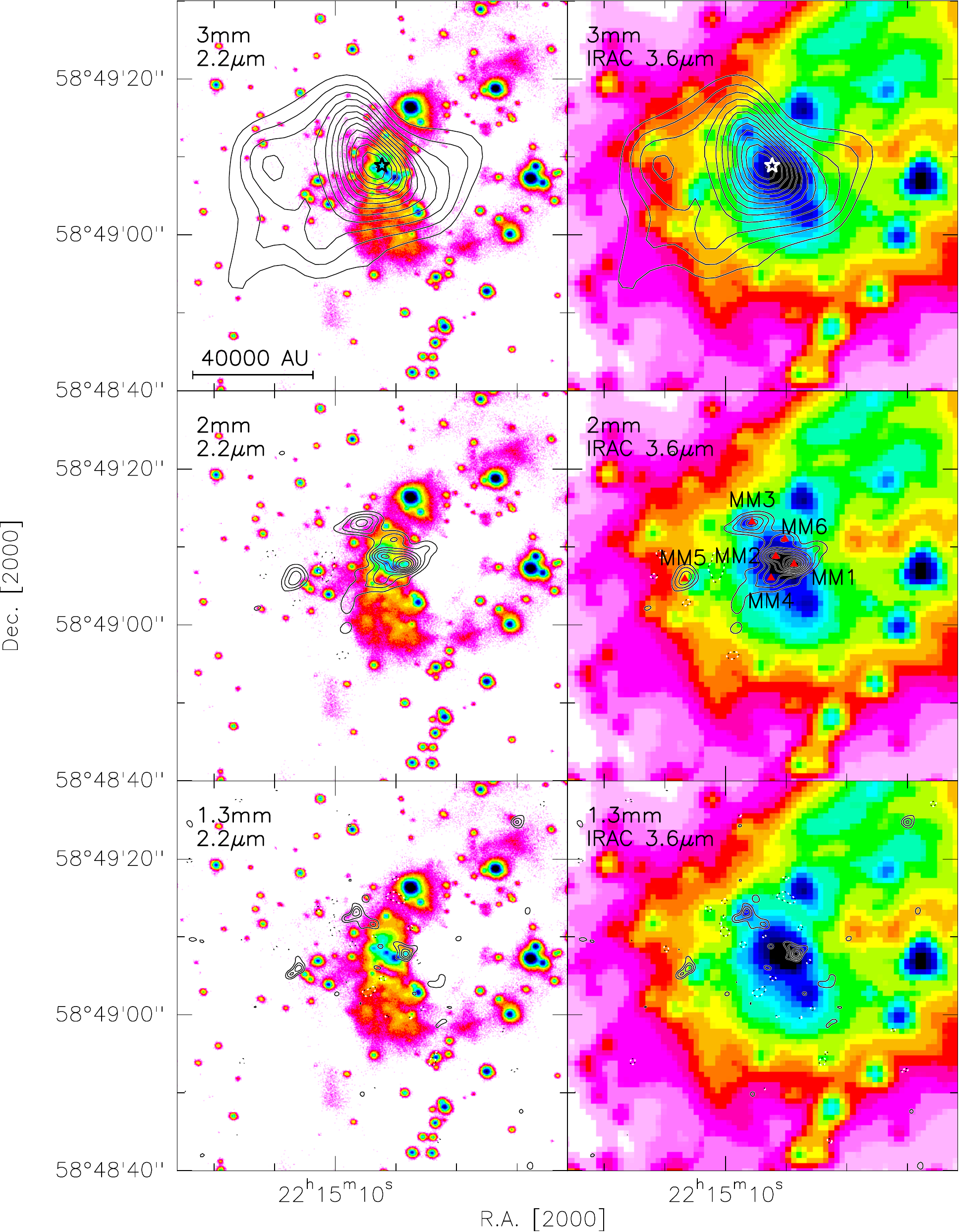}
   \caption{Millimeter continuum map obtained with the CARMA, PdBI, and SMA overlaid with UKIRT $K-$band (2.2~$\mu$m, left) and {\it Spitzer}/IRAC 3.6~$\mu$m (right) images. Contour levels start at 3$\sigma$ in steps of 2$\sigma$ in all panels. The stars in the {top panels} mark the UCH{\sc ii} region VLA1. The UKIRT $K-$band image is from \citet{kumar2003}. The {\it Spitzer}/IRAC post-bcd data processed with pipeline version S18.25.0 have been downloaded from the {\it Spitzer} archive to create these images.}
     \label{fig_ircont}
\end{figure}

Figure \ref{fig_mmcont} shows the millimeter continuum maps of I22134 obtained with CARMA (3~mm), PdBI (2~mm), and SMA (1.3~mm), and Figure~\ref{fig_ircont} shows the continuum maps overlaid on the UKIRT $K-$band (2.2~$\mu$m) and {\it Spitzer}/IRAC 3.6~$\mu$m images. At 3~mm, we detected one main millimeter clump with weak extended emission to the southeast. With much better angular resolution, the clump is resolved into multiple cores at 2~mm and 1.3~mm. In the 3~mm continuum map, the size of the clump is outlined by the 3$\sigma$ contour (1$\sigma$=0.38~mJy~beam$^{-1}$, Fig.~\ref{fig_mmcont}). 

For the 2~mm and 1.3~mm continuum maps, we employed the application FINDCLUMPS\footnote{FINDCLUMPS is a part of the open source software package Starlink, which was a long-running UK Project supporting astronomical data processing. It was shut down in 2005, but the software continued to be developed at the Joint Astronomy Centr until March 2015, and is now maintained by the East Asian Observatory. with the algorithm ClumpFind \citep{williams1994} and 3$\sigma$ boundaries (1$\sigma$=0.26~mJy~beam$^{-1}$ for 2~mm and $=0.50$~mJy~beam$^{-1}$ for 1.3~mm continuum map) to identify the cores. The identified cores that are located outside the FWHM of the primary beam (32$^{\prime\prime}$ for the PdBI, and 55$^{\prime\prime}$ for the SMA) or have a peak intensity less than 5$\sigma$ are rejected manually. Eventually, we identified 6 cores at 2~mm and 5 at 1.3~mm. The properties of the cores we identified are listed in Table~\ref{tab_cont}. The sizes and fluxes are derived from ClumpFind, and the peak intensities and positions are measured from the continuum maps directly.}. Figure \ref{fig_mmcont} shows that the 2~mm core MM1 is associated with one 1.3~mm core, MM3 is resolved into two 1.3~mm cores, named MM3a and MM3b. The 2~mm core MM5 is resolved into two cores at 1.3~mm, named MM5a and MM5b. MM2 is associated with one NIR source seen in the $K-$band image (Fig.~\ref{fig_ircont}) and the UCH{\sc ii} region VLA1 (Fig.~\ref{fig_cmcont}). Figure~\ref{fig_ircont} shows that MM1 coincides with one NIR source in the $K $band. MM3a shows no emission in the $K $band, but strong emission is detected toward this source at longer wavelengths in {\it Spitzer}/IRAC bands (Fig.~\ref{fig_ircont}), which indicates that MM3a is still relatively deeply embedded. 

We did not detect the 1.3~mm counterparts of MM2, MM4, and MM6, which could be due to the 2~mm observations having better uv coverage and/or sensitivity than the 1.3~mm observations. If we assume the dust emissivity index of $\beta=1.8$ and convert the peak intensity of the MM2, MM4, and MM6 we measured at 2~mm (Table~\ref{tab_cont}), the results would be above the 3$\sigma$ detection limit we have with the SMA observations, so the higher sensitivity of the PdBI is not the reason we do not detect the 1.3~mm counterparts of these cores. Following the method described in the appendix of \citet{palau2010}, we estimated that the largest structure from the shortest baselines of our observations that our PdBI observations could recover is $\sim 9.5\arcsec$, while for the SMA observation it is $\sim 5.7\arcsec$. Thus the relatively smooth distributed MM2, MM4, and MM6 were filtered out by the SMA observations. VLA2 is not detected at millimeter wavelengths, probably due to the low sensitivity of our mm observations. Figure~\ref{fig_mmcont} shows the continuum cores detected by \citet{palau2013}, and the southwest one is associated with MM1. We will discuss the detailed comparison in Sect.~\ref{sec_dis_cont}.
-eps-converted-to.pdf

Assuming optically thin dust emission and following Eqs.~\ref{eq_mass} and \ref{eq_dense} \citep{hildebrand1983, schuller2009}, we estimate the gas mass and peak column density of the continuum sources:

\begin{equation} \label{eq_mass}
   \centering
M_{\rm gas}=\frac{R~S_\nu~d^2}{B_\nu (T_{\rm dust})~\kappa_\nu}
\end{equation}

\begin{equation} \label{eq_dense}
   \centering
N_{\rm H_2}=\frac{R~I_\nu}{B_\nu (T_{\rm dust})~\Omega~\kappa_\nu~\mu~m_{\rm H}}
\end{equation}
where $S_\nu$ is the total flux density integrated over the source, $I_\nu$ is the peak intensity of the source (Table~\ref{tab_cont}), $d$  the distance to the source $\sim$2.6~kpc, $R$  the gas-to-dust ratio $\sim$100, $\Omega$  the beam solid angle, $\mu$  the mean molecular weight of the interstellar medium $\sim2.3$, $m_{\rm H}$  the mass of an hydrogen atom, $B_\nu$  the Planck function for a dust temperature $T_{\rm dust}$, and $\kappa_\nu$ is the dust opacity $\sim0.899$~cm$^{2}$~g$^{-1}$ at 1.3~mm (0.41~cm$^{2}$~g$^{-1}$ at 2~mm and 0.18~cm$^{2}$~g$^{-1}$ at 3~mm for the thin ice mantles at the density of 10$^6$~cm$^{3}$, a dust emissivity index of $\beta=1.8$, \citealt{ossenkopf1994}). Assuming local thermodynamic equilibrium (LTE), the rotational temperature ($T_{\rm rot}$) obtained from the NH$_3$ observations (Sect. \ref{sec_dens}) approximates kinetic temperature $T_{\rm k}$. At the densities of our cores ($\>10^5$~cm$^{-3}$), the dust and gas are coupled well, so we take $T_{\rm rot}$ to be equal to dust temperatures $T_{\rm dust}$. 

In cores for which the $T_{\rm rot}$ measurements are not available, a dust temperature of 20~K is assumed. The results are shown in Table~\ref{tab_cont}. Adopting the spectral index $\alpha=-0.19$ we derived for the centimeter emission of VLA1 in Sect.~\ref{sec_cmcont}, we estimate the free-free flux at 2~mm (2.0~mJy) and 3~mm (2.2~mJy) and correct the mass and column density estimations that are listed for MM2 and the CARMA clump in Table~\ref{tab_cont}. We took only the error from the intensity and flux measurement into account to estimate the error for the mass and column density estimations as listed in Table~\ref{tab_cont}. The uncertainty for the $T_{\rm rot}$ measurements is $\sim10-20\%$ \citep{busquet2009}, which would also bring $<30\%$ uncertainty into our mass and column density estimation.

\begin{table*}
\caption{\label{tab_cont}Millimeter continuum peak properties.}
\centering
\begin{tabular}{lcccccccr}
\hline\hline
\noalign{\smallskip}
Cores&R.A. (J2000)&Dec. (J2000)&$I_\nu$\tablefootmark{a}&$S_\nu$\tablefootmark{a}&Mass\tablefootmark{b}&$N_{\rm H_2}$\tablefootmark{b}&$T_{\rm dust}$ & Size\\
        &               &                       &(mJy~beam$^{-1}$)&(mJy)&($M_\odot$)&($\times10^{23}$~cm$^{-2}$)& (K) & arcsec$^2$\\      
\hline
\noalign{\smallskip}
3~mm    &               &                       & & && & \\     
\hline
\noalign{\smallskip}
clump\tablefootmark{c}&22:15:09.30&58:49:08.7& 8.8$\pm$0.38&48.8$\pm$ 10.4&         159.5$\pm$33.9& 2.3$\pm$ 0.1&25\tablefootmark{d}&736.0\\

\hline
\noalign{\smallskip}
2~mm    &               &                       & & & &&  \\    
\hline
\noalign{\smallskip}
MM1&22:15:08.87&58:49:07.8& 4.9$\pm$0.26&9.3$\pm$  2.0& 6.1$\pm$ 1.3& 2.5$\pm$ 0.1&20&20.8\\
MM2\tablefootmark{c}&22:15:09.17&58:49:08.8& 1.5$\pm$0.26&2.9$\pm$  1.1& 1.9$\pm$ 0.7& 0.8$\pm$ 0.1&20&11.0\\
MM3&22:15:09.56&58:49:13.2& 2.6$\pm$0.26&4.7$\pm$  1.1& 2.2$\pm$ 0.5& 0.9$\pm$ 0.1&27\tablefootmark{d}&15.3\\
MM4&22:15:09.25&58:49:06.0& 2.2$\pm$0.26&4.4$\pm$  1.1& 2.9$\pm$ 0.7& 1.1$\pm$ 0.1&20&16.1\\
MM5&22:15:10.67&58:49:05.9& 2.1$\pm$0.26&2.6$\pm$  0.7& 1.7$\pm$ 0.4& 1.1$\pm$ 0.1&20\tablefootmark{d}&8.6\\
MM6&22:15:09.02&58:49:11.0& 1.9$\pm$0.26&2.0$\pm$  0.6& 1.3$\pm$ 0.4& 1.0$\pm$ 0.1&20&7.2\\

\hline
\noalign{\smallskip}
1.3~mm  &               &                       & & & &  &\\    
\hline
\noalign{\smallskip}

MM1&22:15:08.84&58:49:07.8& 5.3$\pm$0.5&9.5$\pm$  2.2& 1.5$\pm$ 0.3& 2.0$\pm$ 0.2&20&4.9\\
MM3a&22:15:09.66&58:49:13.2& 4.5$\pm$0.5&8.4$\pm$  2.0& 0.9$\pm$ 0.2& 1.2$\pm$ 0.1&27\tablefootmark{d}&4.6\\
MM3b&22:15:09.38&58:49:11.6& 2.9$\pm$0.5&4.2$\pm$  1.2& 0.5$\pm$ 0.1& 0.8$\pm$ 0.1&25\tablefootmark{d}&2.9\\
MM5a&22:15:10.62&58:49:06.0& 4.4$\pm$0.5&4.7$\pm$  1.2& 0.7$\pm$ 0.2& 1.7$\pm$ 0.2&20\tablefootmark{d}&2.6\\
MM5b&22:15:10.72&58:49:05.2& 3.7$\pm$0.5&2.2$\pm$  0.7& 0.3$\pm$ 0.1& 1.4$\pm$ 0.2&20\tablefootmark{d}&1.2\\

\hline
\end{tabular}
\tablefoot{
\tablefoottext{a}{The errors are estimated with the same method as described in the note of Table~\ref{tab_vla}, and $\%_{\rm uncertainty}$ equals 20$\%$ for all the mm observations.}\\
\tablefoottext{b}{The error in the mass and column density is estimated by taking only the error in intensity and flux measurements into account. }\\
\tablefoottext{c}{The results for these sources are corrected for free-free emission.} \\
\tablefoottext{d}{$T_{\rm dust}$ for these sources are derived from $T_{\rm rot}$ obtained from the NH$_3$ observations (Sect. \ref{sec_dens}), and for the rest of the sources, $T_{\rm dust}$ is assumed to be 20~K.} \\
}
\end{table*}

\subsection{Molecular line emission}
\label{sec_lines}
\subsubsection{Large scale}
\begin{figure*}[htbp]
   \centering
   \includegraphics[width=0.9\linewidth]{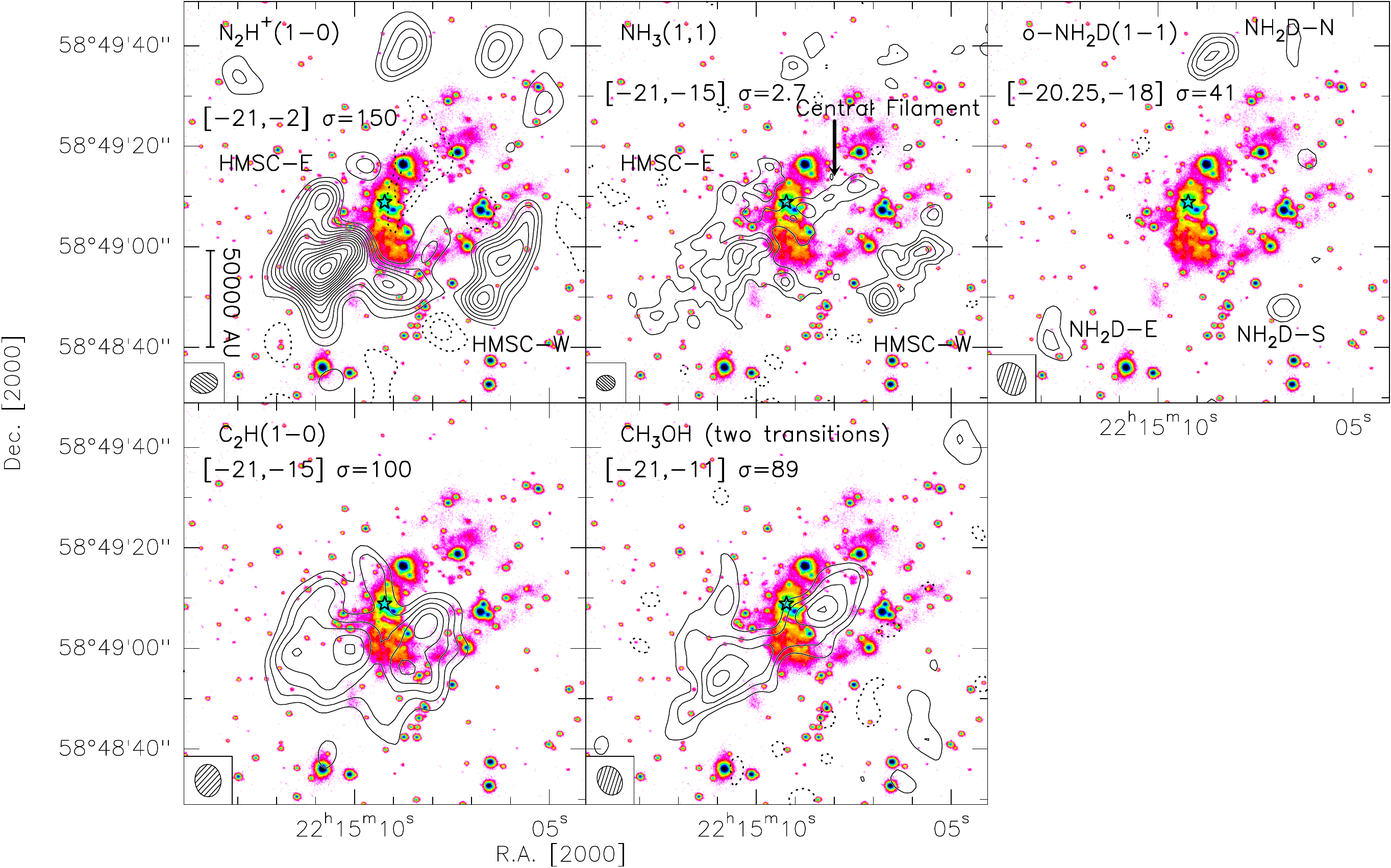}
   \caption{UKIRT $K-$band image and large scale molecular-line-integrated intensity maps. The contour levels start at 3~$\sigma$ and increase with a step of 2~$\sigma$/level in all panels. The $\sigma$ value and the integrated velocity ranges of each transition are shown in the relative map in the units of mJy~beam$^{-1}$ km~s$^{-1}$ and km~s$^{-1}$, respectively. As the integrated velocity ranges show, we include all the hyperfine components N$_2$H$^+$(1--0) for the integrated intensity map, and for NH$_3$(1, 1) we integrate only the main group of hyperfine components. The star indicates the UCH{\sc ii} region, and the crosses the ring-shaped cluster of infrared sources identified by \citet{kumar2003}. The dotted contours are the negative features due to the missing flux with the same contour levels as the positive ones in each panel. The synthesized beam is shown in the bottom left corner of each panel.}
     \label{fig_lines}
\end{figure*}

The CARMA and VLA molecular line observations are used to study the large scale structure of the molecular cloud in this region. Most of the molecular gas is concentrated at the southeast edge of the NIR cluster, and the cometary arc of the H{\sc ii} region VLA1 is also pointing in this direction. Figure 5 shows that the molecular cloud traced by N$_2$H$^+(1-0)$ and NH$_3$(1, 1) consists of several clumps that form an extended structure around the UCHII region and the NIR cluster. Two main structures can be identified in the N$_2$H$^+(1-0)$ and NH$_3$(1, 1) maps, one to the southeast of the NIR cluster and the other one to the southwest. The two main structures are not associated with any embedded infrared source (Fig.~\ref{fig_lines}), thus we name them high-mass starless clumps east (HMSC-E) and west (HMSC-W). 

Several small clumps are detected in N$_2$H$^+(1-0)$ and NH$_3$(1, 1) maps, which are located in the north and northwest of the NIR cluster. Additionally, the NH$_3$(1, 1) emission reveals an elongated filament extending from the southeast to the northwest across the center of the ring shape of the NIR cluster. We call it the central filament (Fig.~\ref{fig_lines}). Three deuterated ammonia {\it ortho}-NH$_2$D clumps are detected surrounding the NIR cluster, and we name them NH$_2$D-N (north), NH$_2$D-E (east), and NH$_2$D-S (south), respectively. NH$_2$D-N is associated with NH$_3$(1, 1) and N$_2$H$^+$(1--0) emission, NH$_2$D-S is associated with the southern peak of HMSC-W, and NH$_2$D-E sits close to the southeastern tail of the HMSC-E. Single-dish observations by \citet{fontani2015} detected {\it ortho}-NH$_2$D emission toward the UCH{\sc ii} region and HMSC-E with the same peak intensity of 0.05~K (corresponding 200~mJy with a beam size of $\sim 28\arcsec$), and this intensity is only one third of the intensity they detected toward HMSC-W. The 3$\sigma$ limit for our {\it ortho}-NH$_2$D observation is $\sim$273~mJy~beam$^{-1}$ (Table~\ref{tab_lines}), and higher than the peak intensity \citet{fontani2015} detected toward the UCH{\sc ii} region and HMSC-E. Strong C$_2$H($N=$1--0) and CH$_3$OH emission is also detected in the southeast of the NIR cluster and extended toward the ``cavity'' of the NIR cluster.

\subsubsection{Small scale}
\label{sec_outflow}
\begin{figure*}[htbp]
   \centering
   \includegraphics[width=0.9\linewidth]{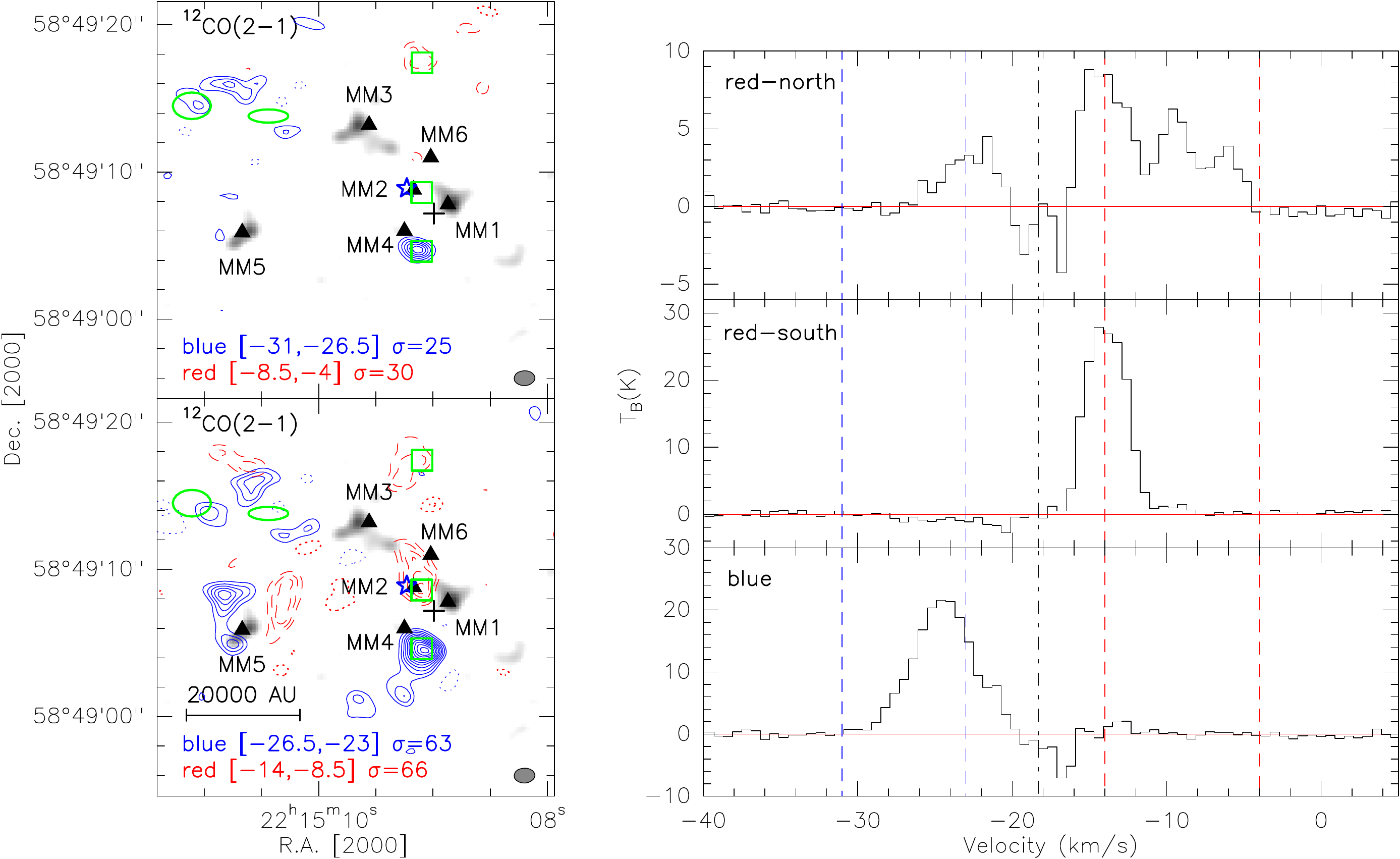}
   \caption{SMA $^{12}$CO(2--1) outflow maps overlaid with the SMA 1.3~mm continuum map ({\it left}), and the average $^{12}$CO(2--1) spectra extracted from the peaks of the red- and blueshifted emission denoted by squares in the {\it left} panels. {\it Top-left} panel shows the high velocity outflow map, and the {\it bottom left} one shows the low-velocity outflow map. In both panels, the full and dashed contours show the blueshifted and redshifted $^{12}$CO(2--1) outflow emission, respectively. All the contour levels start at 4$\sigma$ and increase with a step of 2$\sigma$/level. The $\sigma$ value and velocity-integration range of each outflow emission are shown in the bottom of the left panel in units of mJy~beam$^{-1}$ and km~s$^{-1}$, respectively. Two open ellipses indicate the nebulous H$_2$ emission detected by \citet{kumar2002}. The cross marks the infrared source that was suggested by \citet{palau2013} as the driving source of the outflow. The triangles indicate the positions of our 2~mm cores. The dotted contours are the negative features due to the missing flux with the same contour levels as the positive ones in the panel. The synthesized beam of the outflow map is shown in the bottom right corner of the left panel. The squares at the peaks of the red and blue outflow lobes with the size of 1$\farcs4\times1\farcs4$ show the area where we extracted the spectra shown in the {\it right} panels. In the {\it right} panels, the dashed lines indicate the velocity regimes we used to define the outflow emission, the dashed-dotted line marks the $v_{\rm LSR}=-18.3$~km~s$^{-1}$. }
     \label{fig_out}
\end{figure*}

In addition to the continuum emission, our extended configuration SMA observations also detect the CO(2--1) and SO($6_5-5_4$) lines. The $^{12}$CO(2--1) outflow maps are shown in the lefthand panels of Fig.~\ref{fig_out}. The $^{12}$CO(2--1) spectra (right panels, Fig.~\ref{fig_out}), which were extracted from the related peaks of the red- and blueshifted emission (left panels, Fig.~\ref{fig_out}), show that most of the emission close to the $v_{\rm LSR}$ is filtered out by the interferometer. The $^{12}$CO(2--1) spectrum toward the blueshifted emission peak shows strong blueshifted emission but no redshifted emission. Similarly, the spectrum extracted toward the red-south peak shows strong redshifted emission but no blueshifted emission. The spectrum toward the red-north peak shows weak blueshifted emission at low velocity ({ --26.5} to --23~km~s$^{-1}$) and strong redshifted emission at both high and low velocity. The $^{12}$CO(2--1) outflow maps shows a bipolar outflow that is roughly in the south-north direction. The outflow maps show that the candidate driving source could be MM2, MM4, or the infrared source IRS1 suggested by \citet{palau2013} (see Fig.~3 in \citealt{palau2013}), which is marked in Fig.~\ref{fig_out}. However, it seems less likely that MM2 could be powering such a collimated outflow while being in an advanced evolutionary stage where it has already cleaned up most of the surrounding gas (see Sects.~\ref{sec_bubble} and \ref{sec_pic}). The redshifted emission resembles the Hubble law with the increasing velocity at a longer distance from the driving source \citep{arce2007}. Previous single-dish studies also detect outflow activity with a direction of southeast (blue)-to-northwest (red) in this region (e.g., \citealt{beuther2002b, lopez-sepulcre2010}). With a higher angular resolution PdBI A configuration observations (0\farcs58$\times$0\farcs49), \citet{palau2013} filter out all the large scale extended emission and resolve two outflow knots, one blueshifted in the south and one redshifted in the north with the size of $\sim1-2\arcsec$. { These outflow knots coincide with the main south-north direction blue- and redshifted peak in our outflow map (Fig.~\ref{fig_out}).} NIR narrow band observations by \citet{kumar2002} detected knotty, nebulous H$_2$ emission in the northeast direction of the redshifted outflow component (green ellipses in Fig.~\ref{fig_out}), and they claim the H$_2$ emission is tracing the outflows. The blueshifted emission detected in both outflow maps in the northeastern part of the region could be associated with the H$_2$ emission. It is clearly shown in the outflow map (Fig.~\ref{fig_out}) that the blue- and redshifted components are not lined up straight, which might be due to the inhomogeneous distribution of the ambient molecular gas or to the strong influence of the cluster and nearby stars as suggested by \citet{kumar2002}. { Besides the main south-north outflow}, the outflow maps also show many other red- and blueshifted lobes spread through the image, which could be because this outflow is extended, and we filter out a lot of the large scale structure or multiple outflows driven by different sources. Multi-configuration observations, as well a complementing short spacing single-dish data in order to have a good {\it uv-}coverage, are necessary to understand the outflow emission structure in this region. We combined the low- and high-velocity outflow emission, and estimated the mass $M_{\rm out}$, momentum $p_{\rm out}$ and energy $E_{\rm out}$ of the bipolar outflow, which is in the south-north direction (outflow lobes: squares in Fig.~\ref{fig_out}). For the calculations, we applied the method of \citet{cabrit1990, cabrit1992}, assuming that the $^{13}$CO(2--1)/$^{12}$CO(2--1) line wing ratio is 0.1 \citep{choi1993, beuther2002b}, and H$_2$/$^{13}$CO$=89\times10^4$ \citep{cabrit1992} and $T_{\rm ex}=30$~K. For the blueshifted emission, we derived $M_{\rm out}\sim$0.07~$M_\odot$ and $p_{\rm out}\sim$0.85~$M_\odot$~km~s$^{-1}$, and for the redshifted emission $M_{\rm out}\sim$0.11~$M_\odot$ and $p_{\rm out}\sim$1.52~$M_\odot$~km~s$^{-1}$. The total energy of the outflow $E_{\rm out}\sim3.24\times10^{44}$~erg ({ corresponding to 16.29~$M_\odot$~(km~s$^{-1}$)$^2$}). From single-dish measurements, \citet{beuther2002b} derived  a total outflow mass of $M_{\rm out}=$17~$M_\odot$ total outflow momentum $p_{\rm out}=$242~$M_\odot$~km~s$^{-1}$, and $E_{\rm out}$=3.4$\times10^{46}$~erg, which indicates that we only recovered $\sim$1\% of the outflow emission.

\begin{figure}[htbp]
   \centering
   \includegraphics[width=1\linewidth]{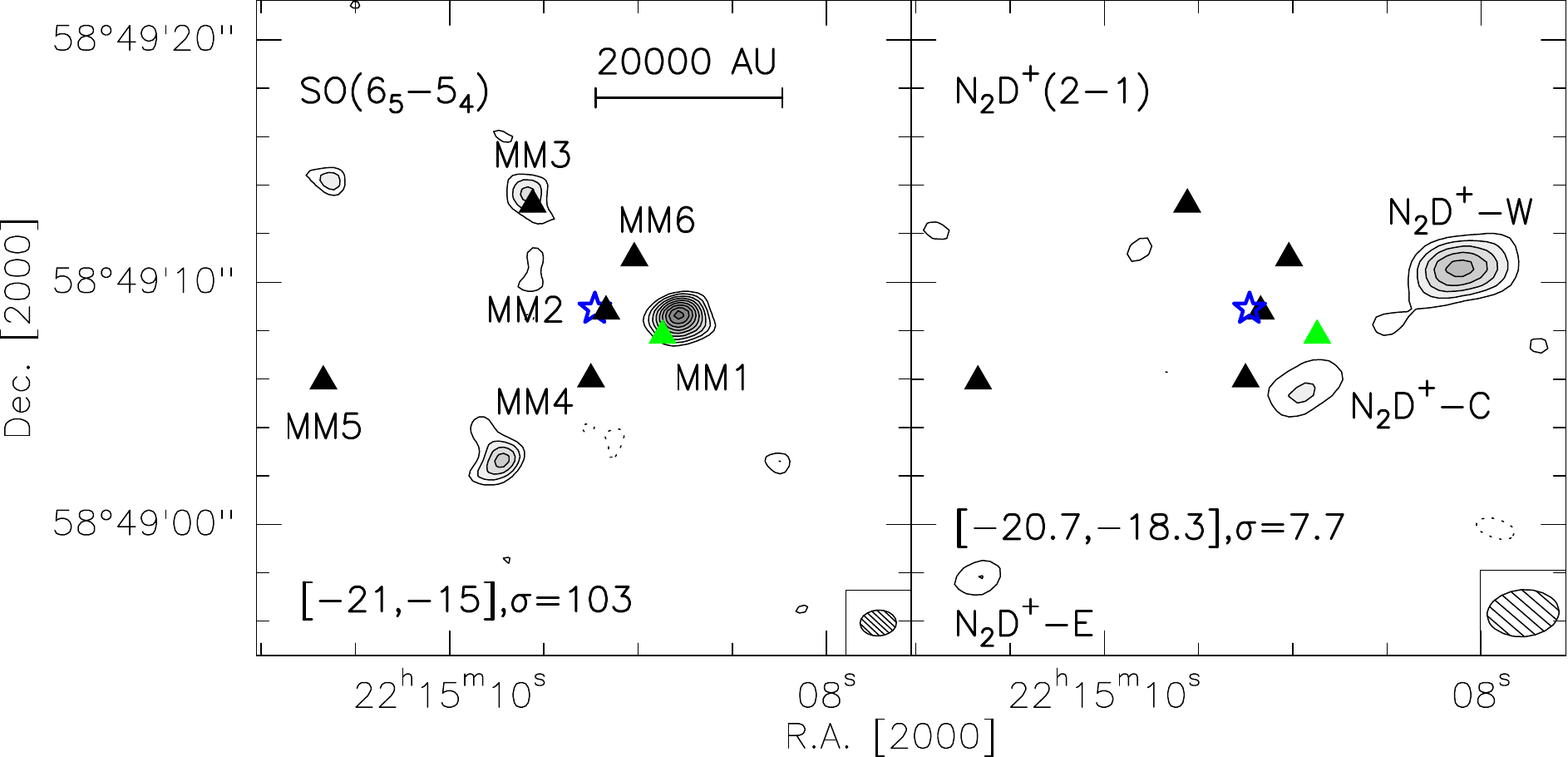}
   \caption{SMA SO(6--5) and PdBI N$_2$D$^+$(2--1) integrated intensity maps. Contour levels start at 3$\sigma$ and increase in steps of 1$\sigma$. The $\sigma$ value and the integrated velocity ranges of each transition are shown in the respect map in the units of mJy~beam$^{-1}$ km~s$^{-1}$ and km~s$^{-1}$, respectively. The star marks the UCH{\sc ii} region and the triangles the 2~mm continuum sources. The dotted contours are the negative features due to the missing flux with the same contour levels as the positive ones in each panel. The synthesized beam is shown in the bottom right corner of each panel.}
     \label{fig_son2dp}
\end{figure}

The integrated intensity map of SO($6_5-5_4$) (left panel, Fig.~\ref{fig_son2dp}) shows one emission peak at the continuum source MM1 and another peak in the position of R.A. $\sim$22$^{\rm h}$15$^{\rm m}$09$^{\rm s}$.7 Dec. $\sim$+58\degr49\arcmin03$\arcsec$ (J2000), which could be tracing part of the large scale molecular cloud surrounding the NIR cluster (Fig.~\ref{fig_lines}),  and is mostly filtered out by the extended configuration SMA observations.

\begin{figure}[htbp]
   \centering
   \includegraphics[width=1\linewidth]{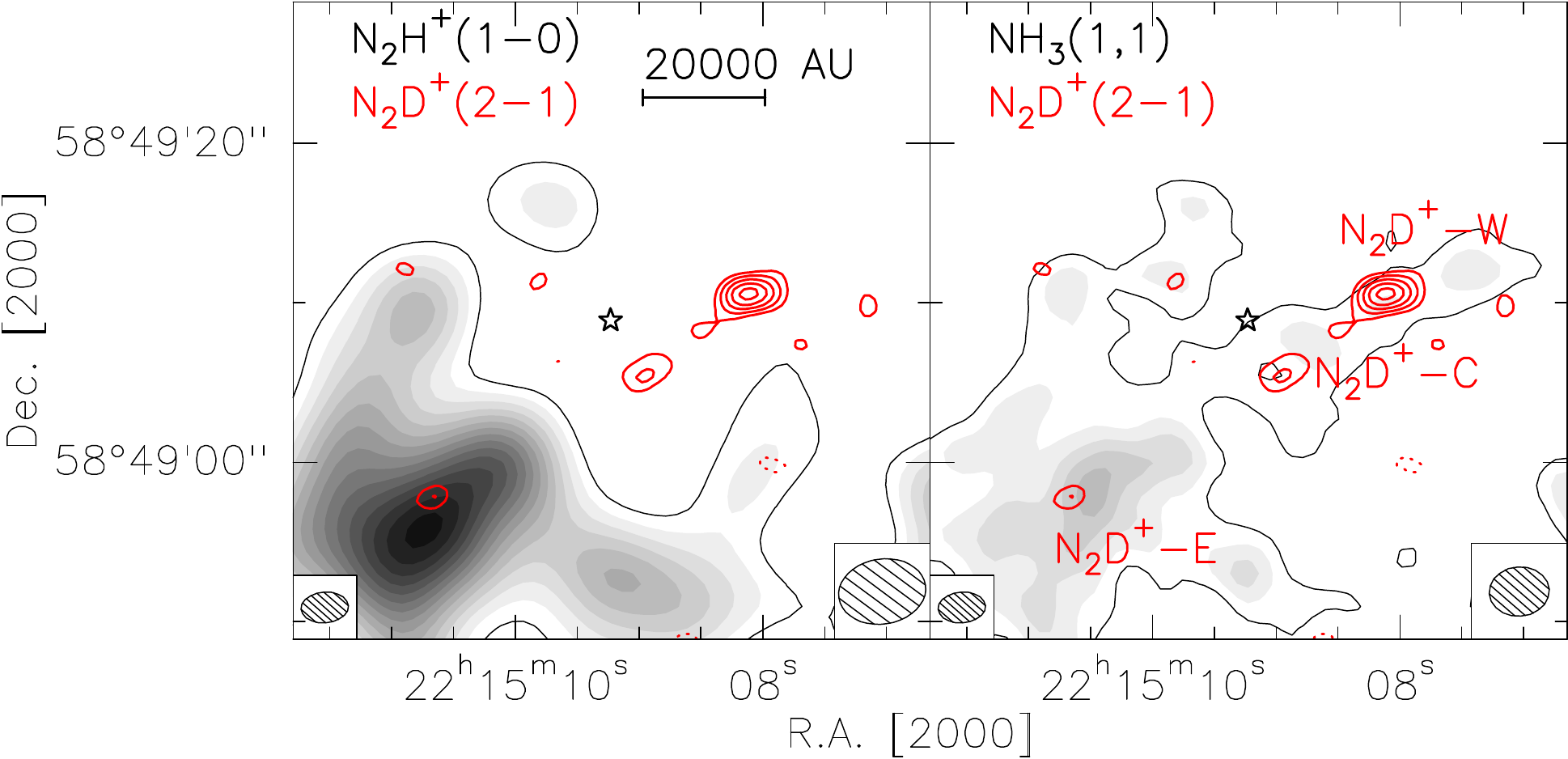}
   \caption{N$_2$D$^+$(2--1) integrated intensity map overlaid on the N$_2$H$^+$(1--0) and NH$_3$(1, 1) integrated intensity maps. The contour level parameters of N$_2$D$^+$(2--1) integrated intensity map are the same as in Fig.~\ref{fig_son2dp}. The thin black contour in each panel indicates the 3$\sigma$ level of the N$_2$H$^+$(1--0) and NH$_3$(1, 1) integrated intensity emission. The $\sigma$ values and integrated velocity range are the same as in Figs.~\ref{fig_son2dp} and \ref{fig_lines}, respectively. The star marks the UCH{\sc ii} region. The dotted contours are the negative features due to the missing flux with the same contour levels as the positive ones in each panel. The synthesized beam of N$_2$D$^+$(2--1) is shown in the bottom left corner of each panel, and the one of N$_2$H$^+$(1--0) and NH$_3$(1, 1) is shown in the bottom right corner of \textcolor{red}{respective} panel.}
     \label{fig_n2dp}
\end{figure}

Figures~\ref{fig_son2dp} and \ref{fig_n2dp} show the integrated intensity map of N$_2$D$^+$(2--1). Three N$_2$D$^+$ cores are detected with a peak intensity $\geq4~\sigma$. We name these cores N$_2$D$^+$-E (east), N$_2$D$^+$-C (center), and N$_2$D$^+$-W (west). N$_2$D$^+$-E is associated with HMSC-E with strong N$_2$H$^+$ emission, and the other two N$_2$D$^+$ cores lie very close to the continuum cores and the UCH{\sc ii} region VLA1 and have no N$_2$H$^+$ emission (left panel, Fig.~\ref{fig_n2dp}). It is worth noting that N$_2$D$^+$-C and N$_2$D$^+$-W lie just between the NH$_3$ peaks, this anticorrelation may indicate some peculiar chemical process. The non-detection of N$_2$H$^+$ emission toward the other two N$_2$D$^+$ cores might be due to a problem with sensitivity, the rms of N$_2$D$^+$ observations being about 20$\%$ of the rms for N$_2$H$^+$ observations (Table~\ref{tab_lines}). The NH$_3$(1, 1) filament that the N$_2$D$^+$ cores are associated with is observed with the VLA  and also has much higher sensitivity and uv-coverage than the N$_2$H$^+$ observations (Table~\ref{tab_lines}). Still, these are the closest N$_2$D$^+$ cores to an UCH{\sc ii} region detected so far (projected distance $\sim$8~000~au at $d=2.6$~kpc). With a primary beam size of $\sim32$\arcsec at 2~mm, our PdBI observations can only cover HMSC-E and the UCH{\sc ii} region but not HMSC-W, so we do not have N$_2$D$^+$ for HMSC-W.

\subsection{Analysis}

\subsubsection{Column density and temperature of NH$_3$ and N$_2$H$^+$}
\label{sec_dens}

\begin{figure*}[htbp]
   \centering
   \includegraphics[width=0.9\linewidth]{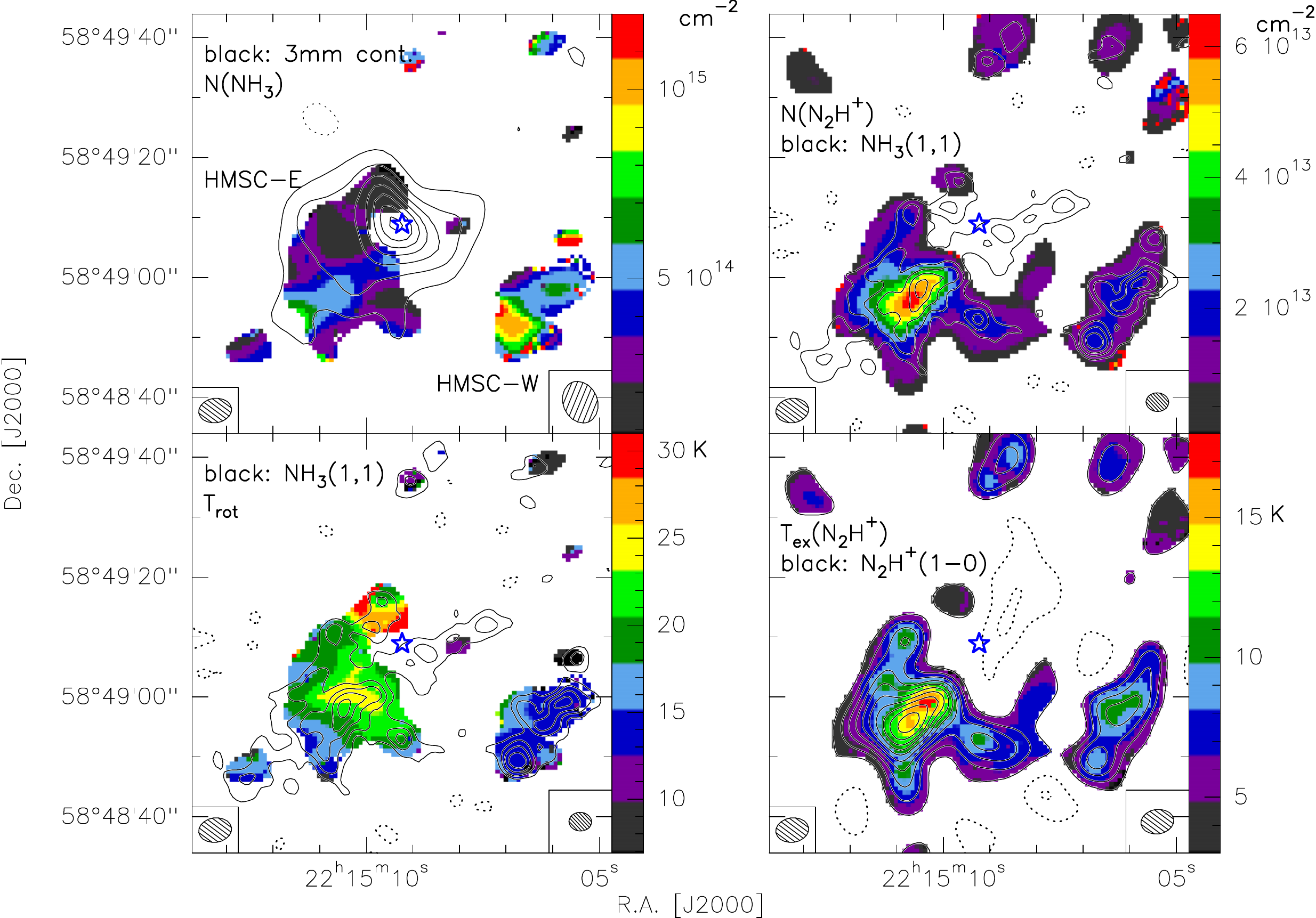}
   \caption{{\it Left:} NH$_3$ column density map ({\it top}) and rotational temperature map overlaid with 3~mm continuum and NH$_3$(1, 1) integrated intensity contours, respectively. {\it Right:} the N$_2$H$^+$ column density map overlaid with NH$_3$(1, 1) integrated intensity contours ({\it top}), and N$_2$H$^+$(1--0) excitation temperature map overlaid with N$_2$H$^+$(1--0) integrated intensity contours. The 3~mm continuum contours start at 3$\sigma$ in steps of 4$\sigma$. All the other contour levels parameters are the same as in Fig.~\ref{fig_lines}. The synthesized beam of the temperature and column density map is shown in the bottom left corner of each panel, and the synthesized beam of the contour map is shown in the bottom right corner of each panel.}
     \label{fig_col}
\end{figure*}

NH$_3$(1, 1), (2, 2), and N$_2$H$^+$(1--0) emission are used to study the column density and temperature properties of the molecular cloud clumps.  To compare with the N$_2$H$^+$ results, we first convolved the NH$_3$(1, 1) and (2, 2) data cube to the same angular resolution as the N$_2$H$^+$(1--0) one, $5\farcs49\times4\farcs09$, P.A.$=-82\degr$. Then we extracted the spectra of NH$_3$(1, 1) and (2, 2) lines for positions in a grid of 0\farcs7$\times$0\farcs7 and fitted the hyperfine structure of the  NH$_3$(1, 1) transition and a Gaussian profile to the NH$_3$(2, 2) using CLASS. Following the procedure described in \citet{busquet2009}, we derived the rotational temperature and the column density maps of the NH$_3$ emission. For N$_2$H$^+$(1--0) transition, we also extracted spectra toward each position and fitted the hyperfine structure of each line \citep{caselli1995}. Following the procedure outlined in the appendix in \citet{caselli2002}, we derived the excitation temperature and column density maps of N$_2$H$^+$ emission. The parameters of the molecule for the calculations are obtained from CDMS \citep{muller2001, muller2005}. The resulting maps are shown in Fig.~\ref{fig_col}. While the peak NH$_3$ column density for HMSC-E is $\sim5\times10^{14}$~cm$^{-2}$, the gas close to the UCH{\sc ii} region shows a lower NH$_3$ column density of $\sim1\times10^{14}$~cm$^{-2}$. The NH$_3$ column density reaches its highest value in the western clump HMSC-W of  $\sim1\times10^{15}$~cm$^{-2}$. In contrast to the NH$_3$ emission, the column density of N$_2$H$^+$ peaks close to the NH$_3$(1, 1) integrated intensity peak with a value of $\sim6\times10^{13}$~cm$^{-2}$, and the western clump HMSC-W only has a peak column density of $\sim2\times10^{13}$~cm$^{-2}$. The $T_{\rm rot}$ map derived from NH$_3$ observations shows that the main part of HMSC-E has a temperature of $\sim$20 to 25~K, which achieves its highest value of 30~K at the position close to the UCH{\sc ii} region. The western clump HMSC-W has a much lower $T_{\rm rot}$ of $\sim14$~K. These results are consistent with the values reported in \citet{sanchez-monge2013}, where the temperature increases when approaching the UV-radiation source. For the N$_2$H$^+$ observations, the excitation temperature $T_{\rm ex}$ also shows a higher temperature at a position close to the UCH{\sc ii} region of $\sim17$~K and a lower value toward HMSC-W of $\sim10$~K.

The deuterated fraction (hereafter $D_{\rm frac}$) of N$_2$H$^+$, defined as the column density ratio of the species containing deuterium to its counterpart containing hydrogen, is reported to be anti-correlated with the evolutionary stages of the high-mass cores \citep{fontani2011}. Following the procedure outlined in the appendix in \citet{caselli2002}, we also estimated the column density of N$_2$D$^+$ and NH$_2$D, and then derived $D_{\rm frac}$(N$_2$H$^+$) and $D_{\rm frac}$(NH$_3$). The results of $D_{\rm frac}$(N$_2$H$^+$) and $D_{\rm frac}$(NH$_3$) are listed in Table~\ref{tab_dfrac}. The N$_2$H$^+$ 3$\sigma$ upper limit was used to estimate $D_{\rm frac}$(N$_2$H$^+$) of the two N$_2$D$^+$ cores close to the UCH{\sc ii}. For simplicity, we assumed $T_{\rm ex}=$20~K and a beam filling factor of 1. Every 10~K change from $T_{\rm ex}$ would add $<22\%$ to our estimation for $D_{\rm frac}$.

\begin{table}
\caption{\label{tab_dfrac} Deuterium fraction of N$_2$H$^+$ and NH$_3$.}
\centering
\begin{tabular}{lc}
\hline\hline
\noalign{\smallskip}
Cores& $D_{\rm frac}$\\
\hline
\noalign{\smallskip}
N$_2$D$^+$-E&   0.0035\\
N$_2$D$^+$-C&   0.035\tablefootmark{$\ast$}\\
N$_2$D$^+$-W&   0.048\tablefootmark{$\ast$}\\
NH$_2$D-E       & 0.040\\
NH$_2$D-S       & 0.012\\
NH$_2$D-N       & 0.014\\
\hline
\end{tabular}
\tablefoot{
\tablefoottext{$\ast$}{Lower limit.} 
}
\end{table}

\subsubsection{Kinematics}
\label{sec_kin}
To study the kinematic properties of the molecular cloud clumps, we constructed velocity maps of C$_2$H($N=$1--0), CH$_3$OH($2_{0, 2}-1_{0, 1})$A+, NH$_3$(1, 1) and N$_2$H$^+$(1--0) with the same spatial and velocity resolution. We convolved all the data cubes to the same spatial (6\farcs49$\times$5\farcs23, P.A.$=22.21$\degr) and velocity resolution (0.42~km~s$^{-1}$) in order to compare the velocity structure among these tracers. A single velocity-component, hyperfine-structure fitting for NH$_3$(1, 1) \citep{busquet2009} and N$_2$H$^+$(1--0) \citep{caselli1995} were performed toward each point and produced the velocity map. For CH$_3$OH, the emission close to the UCH{\sc ii} also shows a double-peaked velocity structure, and the rest shows only one velocity component. The signal-to-noise ratio of the CH$_3$OH($2_{0, 2}-1_{0, 1})$A+ is not good enough to perform the multicomponent fit, so we performed one velocity Gaussian fitting. We found that the C$_2$H($N=$1--0) spectrum shows strong double velocity peaks in some positions, so we followed the routine described by \citet{henshaw2014} and \citet{hacar2013} by performing a multiple-component Gaussian fit. 

\begin{figure*}[htbp]
   \centering
   \includegraphics[width=0.9\linewidth]{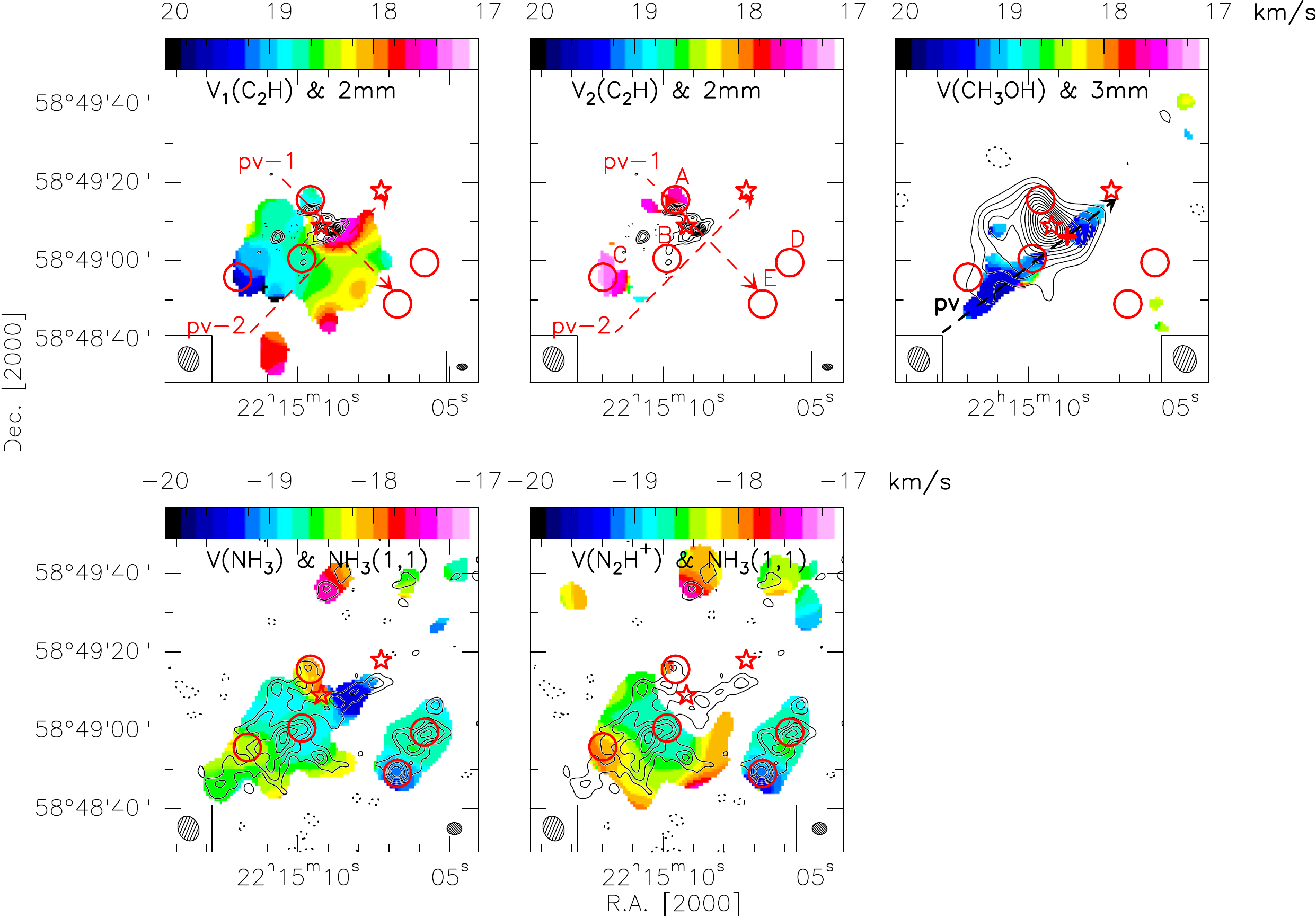}
   \caption{Molecular line velocity maps overlaid with 2~mm, 3~mm continuum map and the NH$_3$(1, 1) integrated intensity map. The contour level parameters are the same as the ones in Figs.~\ref{fig_mmcont} and \ref{fig_lines}. The dashed arrows in the {\it top} panels mark the PV-cuts used for PV diagrams in Fig.~\ref{fig_pv}. The star marks the UCH{\sc ii} region. The cross in the {top right} panel indicateks the reference position in CH$_3$OH PV diagram in Fig.~\ref{fig_pv}. The circles indicate the area where the average spectra are extracted and shown in Fig.~\ref{fig_chspec}. The synthesized beam of the velocity maps is shown in the bottom left corner of each panel, and the one of the contours is shown in the bottom right corner of \textcolor{red}{respect} panel.}
     \label{fig_vel}
\end{figure*}

The process is as follows: We first compute an average spectrum in a square area with a size of $6\farcs49$ (about the size of the synthesized beam) and decide whether the average spectrum has to be fitted with one or two velocity components. The fit results from the average spectrum are used as the initial guesses for all the spectra inside the square area. If one spectrum is fitted with two velocity components, we apply the following quality check: the separation of the two components must be larger than the half-width at half-maximum of the stronger component, and the peak intensity of each component must be larger than 3~$\sigma$ (with $\sigma=$0.45~Jy~beam$^{-1}$), otherwise, the spectrum is fitted with one velocity component (see \citealt{henshaw2014,hacar2013} for the details of the fitting routine). While most of the C$_2$H emission can be fitted with one single velocity component, emission in the north and southwest requires of two velocity components. The blueshifted component shows similar velocity to its adjacent emission, so it is associated with the main emission structure. We combined the velocity map of the blueshifted component with the one of the main emission, and the resulting velocity map is shown in the top lefthand panel in Fig.~\ref{fig_vel} ($V_1$(C$_2$H) ). The velocity map of the redshifted component is shown in panel $V_2$(C$_2$H) in Fig.~\ref{fig_vel}. 

All the velocity maps in Fig.~\ref{fig_vel} are presented with the same velocity range and were clipped at the 4~$\sigma$ level of the respective line channel map. The $V_1$(C$_2$H) velocity map shows a clear velocity gradient, suggesting that the region close to the UCH{\sc ii} region and the NIR cluster is more redshifted than the region far away from the UCH{\sc ii} region. The CH$_3$OH($2_{0, 2}-1_{0, 1})$A+ velocity map shows a more complicated velocity gradient along the filament, where the velocity of the structure shifts from $-19.6$~km~s$^{-1}$ to $-18.9$~km~s$^{-1}$ and back to $-19.6$~km~s$^{-1}$, again following the arrow in Fig.~\ref{fig_vel}. The whole CH$_3$OH($2_{0, 2}-1_{0, 1})$A+ emission structure is also blueshifted by $\sim$1~km~s$^{-1}$ compared to other molecular line emission. Single velocity-component hyperfine-structure fitting for NH$_3$(1, 1) \citep{busquet2009} and N$_2$H$^+$(1--0) \citep{caselli1995} are performed toward each point and produce the velocity map.

The NH$_3$(1,1) velocity map shows that the majority of the emission is at $\sim-18.7$~km~s$^{-1}$. Interestingly, the central filament in the NH$_3$(1, 1) velocity shows blueshifted emission from the main component by $\sim1$~km~s$^{-1}$, while the southern and eastern regions of HMSC-E shows redshifted emission by $\sim1$~km~s$^{-1}$. It is interesting to notice that the N$_2$D$^+$ cores show similar velocity $\sim-19.3$~km~s$^{-1}$ (Fig.~\ref{fig_spec}), which further confirms that these N$_2$D$^+$ cores are associated with the NH$_3$(1, 1) filament. This velocity structure of NH$_3$(1, 1) is also reported by \citet{sanchez-monge2013}. HMSC-E also shows similar velocity structure in the N$_2$H$^+$(1--0) velocity map, where the southeastern part is redshifted and the emission peak is blueshifted. 

\begin{figure*}[htbp]
   \centering
   \includegraphics[width=0.9\linewidth]{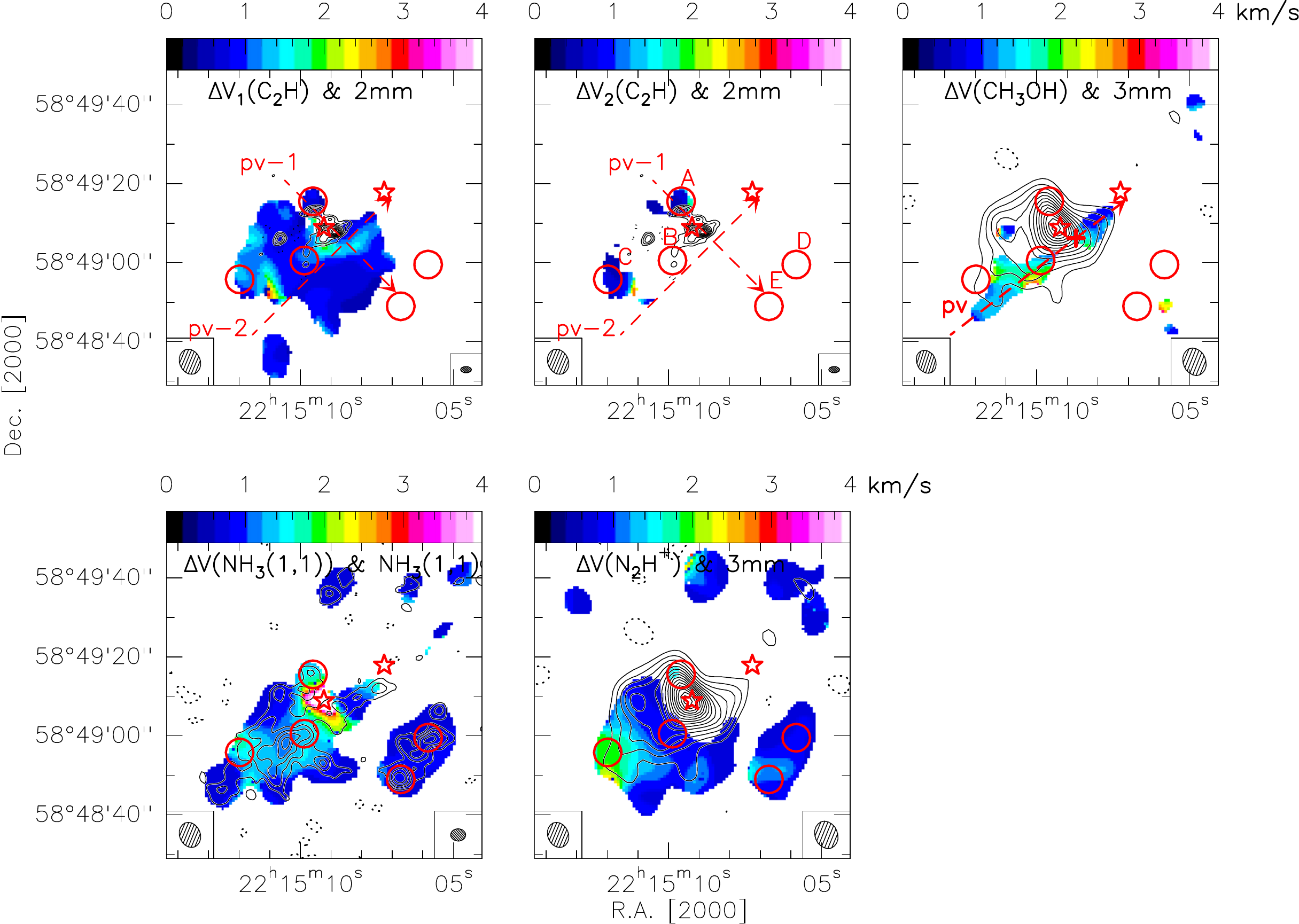}
   \caption{Molecular line-width maps overlaid with the 2~mm, 3~mm continuum map and NH$_3$(1, 1) integrated intensity map. The contour level parameters are the same as the ones in Figs.~\ref{fig_mmcont} and \ref{fig_lines}. The dashed arrows in the {\it top} panels point to the PV-cuts used for PV diagrams in Fig.~\ref{fig_pv}. The star indicates the UCH{\sc ii} region. The cross in the {top right} panel marks the reference position in CH$_3$OH PV diagram in Fig.~\ref{fig_pv}. The circles show  where the average spectra are extracted and shown in Fig.~\ref{fig_chspec}. The synthesized beam of the line-width maps is shown in the bottom left corner of each panel, and the one of the contours is shown in the bottom right corner of \textcolor{red}{respect }panel.}
     \label{fig_width}
\end{figure*}

\begin{figure*}[htbp]
   \centering
   \includegraphics[width=0.9\linewidth]{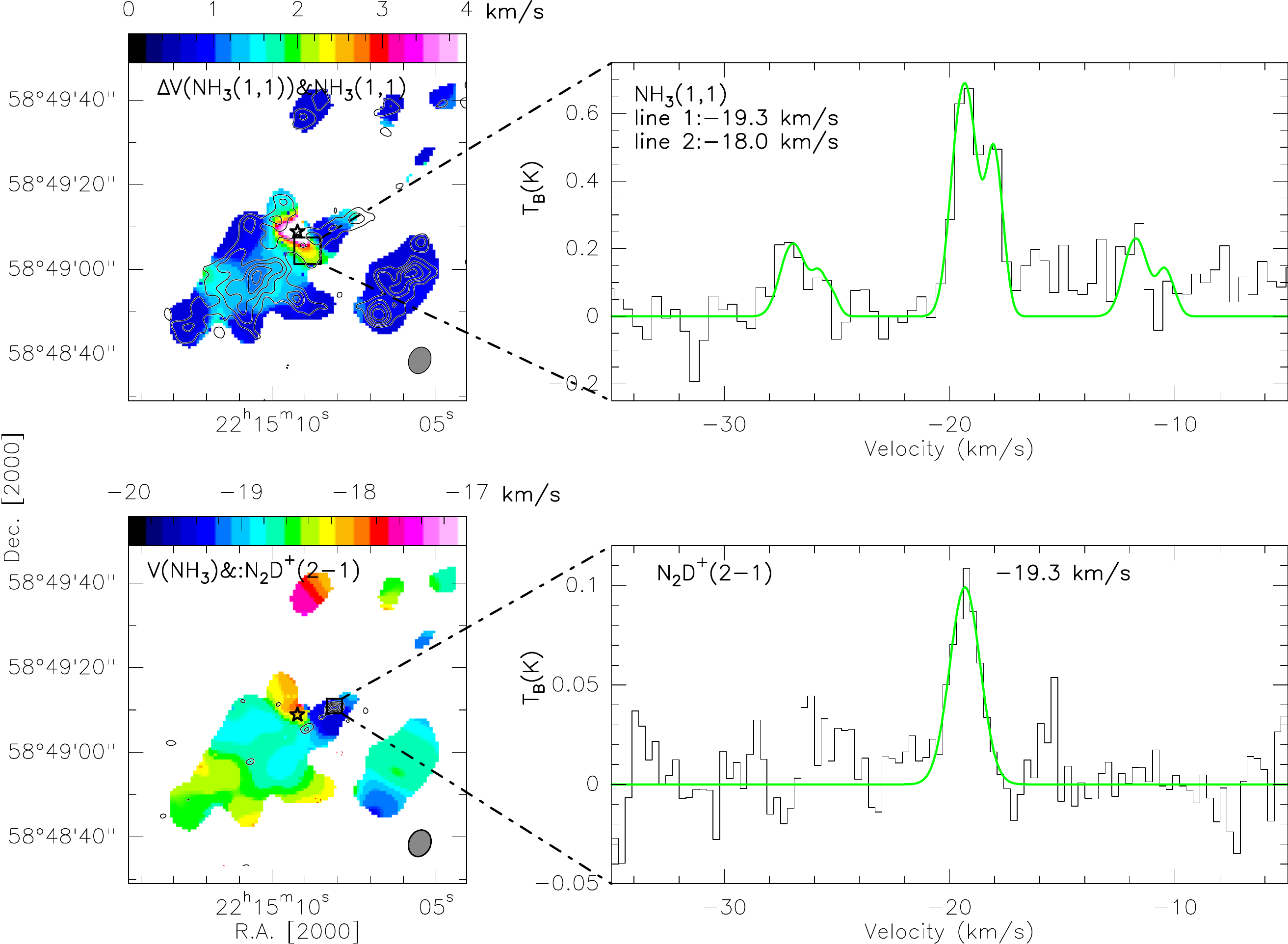}
   \caption{{\it Top:} NH$_3$(1, 1) line-width map overlaid with the NH$_3$(1, 1) integrated intensity map and the average NH$_3$(1, 1) spectrum extracted from the region outlined with the box in the line-width map. {\it Bottom:} NH$_3$(1, 1) velocity map overlaid with N$_2$D$^+$(2--1) integrated intensity map, N$_2$D$^+$(2--1) average spectrum extracted toward the emission peak indicated with the box in the velocity map. }
     \label{fig_spec}
\end{figure*}

The line-width maps of C$_2$H($N=$1--0), CH$_3$OH($2_{0, 2}-1_{0, 1})$A+, NH$_3$(1, 1) and N$_2$H$^+$(1--0) are shown in Fig.~\ref{fig_width}. The line-width maps of NH$_3$(1, 1) and N$_2$H$^+$(1--0) both show that the line width toward western clump HMSC-W is $\sim$0.4~km~s$^{-1}$ smaller than the eastern one HMSC-E, and N$_2$H$^+$(1--0) also shows a large line width ($\sim2$~km~s$^{-1}$) to the southeast of HMSC-E. For  C$_2$H($N=$1--0) and NH$_3$(1, 1), most of the emission shows line widths around 1~km~s$^{-1}$, while close to the UCH{\sc ii} region, the line widths get larger than 2~km~s$^{-1}$, which might be due to the feedback effects from the UCH{\sc ii} region. The NH$_3$(1, 1) spectrum extracted from the joint region (approximately the size of one beam) between the central filament and the emission clump shows two velocity components, one at $\sim$--18.0~km~s$^{-1}$ and the other at $\sim$--19.3~km~s$^{-1}$ (Fig.~\ref{fig_spec}). Furthermore, the central filament has only one velocity component and is $\sim$1~km~s$^{-1}$ blueshifted from the main clump.These features together indicate that the Central Filament is on a slightly different plane from the main clump and the NIR cluster in space. Since the N$_2$D$^+$ cores are associated with the NH$_3$(1, 1) central filament, the proximity of these deuterated cores to the UCH{\sc ii} region could just be a projection effect. Furthermore, the center filament could also produce the central cavity seen in the NIR image (\citealt{kumar2003}, see also Fig.~\ref{fig_lines}) via absorbing the background emission because it is in the foreground of the cluster.

\begin{figure*}[htbp]
   \centering
   \includegraphics[width=0.9\linewidth]{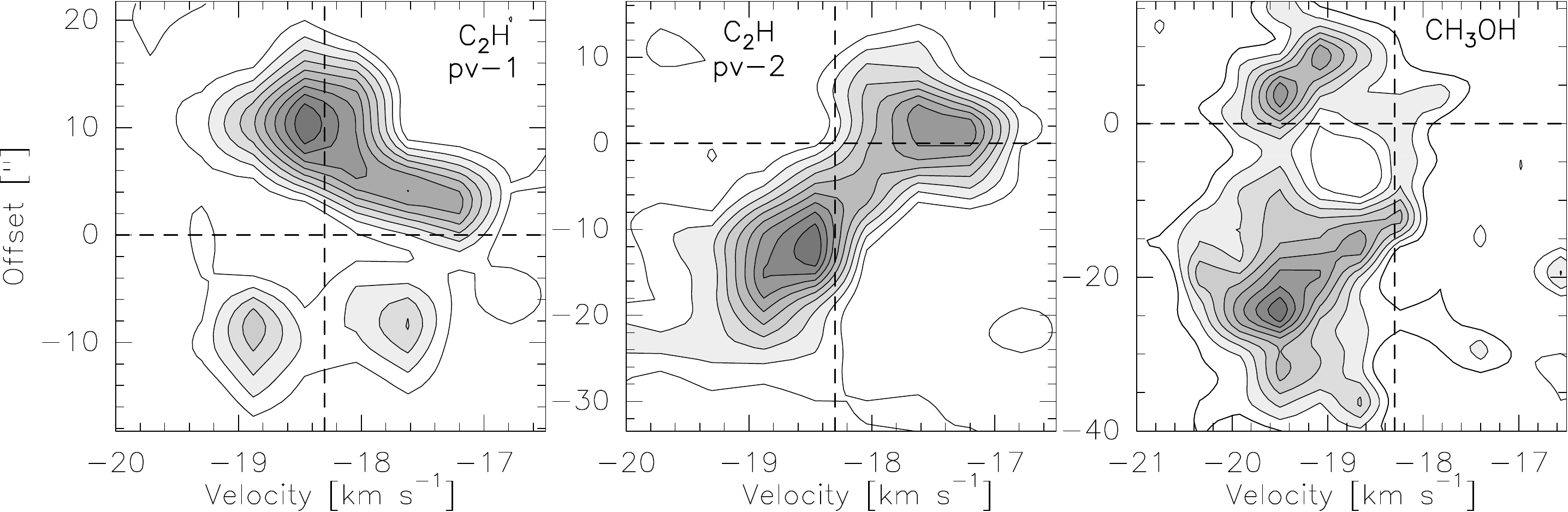}
   \caption{Position-velocity (PV) diagrams of C$_2$H and CH$_3$OH with a velocity resolution of 0.42~km~s$^{-1}$. The PV cuts are shown in the \textcolor{red}{respect} velocity map (Fig.~\ref{fig_vel}). The contour levels are from 10\% to 90\% from the peak emission (from left to right, 0.74, 0.69, and 0.88~Jy~beam$^{-1}$, respectively) in steps of 10\%. The offsets refer to the distance along the cuts from the reference position, which is, from left to right, the dust continuum peak of MM1, the cross position of pv-1 and pv-2 in Fig.~\ref{fig_vel}, and the cross in the CH$_3$OH panel in Fig.~\ref{fig_vel}, respectively. The $v_{\rm LSR}$ at --18.3~km~s$^{-1}$ and the reference position are indicated by vertical and horizontal dashed lines. }
     \label{fig_pv}
\end{figure*}

Figure~\ref{fig_pv} shows the position-velocity (PV) diagrams for C$_2$H($N=$1--0) and CH$_3$OH($2_{0, 2}-1_{0, 1})$A+. The corresponding PV cuts of each panel are shown in Fig.~\ref{fig_vel}. The PV cut (``pv-1'' in the C$_2$H($N=$1--0) velocity map) cuts through the two dust continuum cores MM1 and MM3 (Fig.~\ref{fig_vel}). The PV-diagram shows a C-shape velocity structure, which indicates that the C$_2$H emission is tracing an expanding bubble or shell structure \citep{arce2011}. The PV cut (``pv-2'' in the  C$_2$H($N=$1--0) velocity map in Fig.~\ref{fig_vel}) is perpendicular to the PV cut pv-1. The PV-diagram shows a clear velocity gradient with increasing velocity as we move along the PV cut pv-2, which might be indicating that the emission structure is expanding toward the observer or rotating. As C$_2$H is often considered as a tracer of photo-dominated regions (PDR, e.g., \citealt{millar1984, jansen1995}), the expanding velocity structures might be tracing the expanding PDR driven by the stellar cluster. The PV cut of CH$_3$OH($2_{0, 2}-1_{0, 1})$A+ follows the CH$_3$OH filament (Fig.~\ref{fig_vel}). The PV diagram also shows a ring-shaped velocity structure with the center near the UCH{\sc ii} region (offset 0), which is expected if the emission structure is expanding \citep{arce2011}. CH$_3$OH is well known as a shock tracer (e.g., \citealt{beuther2005a, jorgensen2004}), and the expanding velocity structure could be tracing the interaction front of the molecular cloud and the stellar wind from the cluster members.

\begin{figure*}[htbp]
   \centering
   \includegraphics[width=0.9\linewidth]{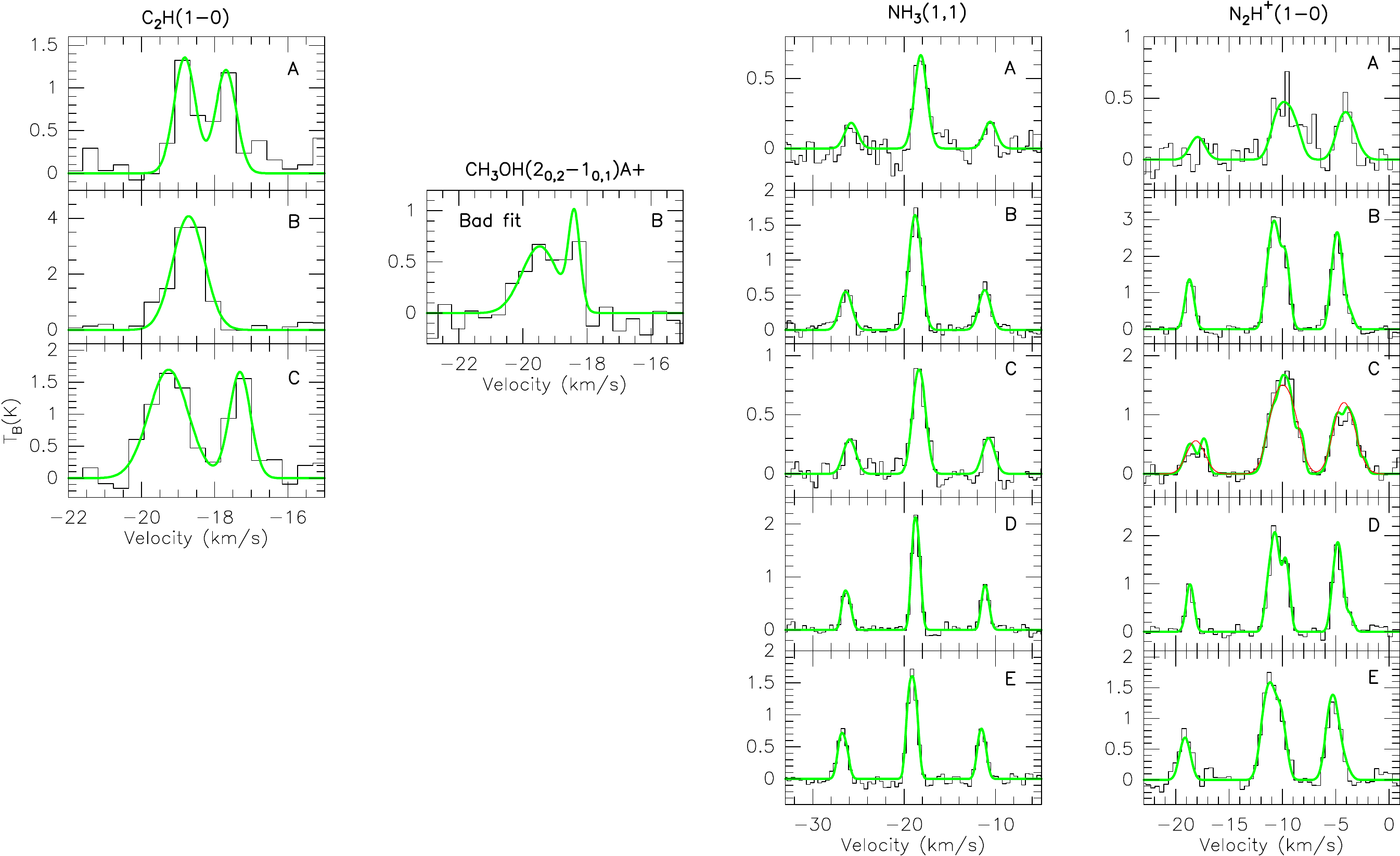}
   \caption{Extracted spectra from the regions marked in Figs.~\ref{fig_vel} and \ref{fig_width}. The thick lines show the fitted profile of the spectra.  Gaussian fittings were performed for CH$_3$OH spectra and C$_2$H($N=$1--0) spectra at positions A and C with two velocity components. The spectrum of C$_2$H($N=$1--0) at position B is fit with one velocity component. Single velocity component hyperfine structure profile fittings are done for all NH$_3$(1, 1) spectra and N$_2$H$^+$(1--0) spectra except the one from position C, in which the thick green line shows the two velocity component fitting and the thin line red shows the one component fitting. { The velocity of N$_2$H$^+$(1--0) spectra is offset by $\sim8$~km~s$^{-1}$ because the rest frequency was set at the frequency of the isolated hyperfine component ($F_1, F=0,1\rightarrow1,2$)}. The results from the fittings are listed in Table~\ref{tab_linefit}.}
\label{fig_chspec}
\end{figure*}

We selected five typical positions, indicated in Fig.~\ref{fig_vel} , and extracted the spectra of C$_2$H($N=$1--0), CH$_3$OH($2_{0, 2}-1_{0, 1})$A+, NH$_3$(1, 1) and N$_2$H$^+$(1--0) to compare the kinematic properties in different locations. Position A was chosen because it shows two velocity components in C$_2$H($N=$1--0) and a large line width in NH$_3$(1, 1) and N$_2$H$^+$(1--0). Position B is close to the emission peak of NH$_3$(1, 1) and also shows two velocity components in CH$_3$OH($2_{0, 2}-1_{0, 1})$A+. Position C shows two velocity components in C$_2$H($N=$1--0) and a large line width in N$_2$H$^+$(1--0). Positions D and E are the peak positions of N$_2$H$^+$(1--0) integrated emission toward HMSC-W. We computed the average spectra within the circular area with a radius of 3\farcs5 (the synthesized is  6\farcs49$\times$5\farcs23) at each position, then single or two velocity components are fitted to the average spectra. The spectra and fitting results are shown in Fig.~\ref{fig_chspec} and Table~\ref{tab_linefit}. In Table~\ref{tab_linefit}, for positions that the line was fit with two velocity components, the redshifted one is labeled with subscript 1 (e.g., A$_1$), and the redshifted one with subscript 2 (e.g., A$_2$). 

While the spectra of C$_2$H($N=$1--0) at positions A and C show clear double-peaked features, the spectrum at position B has only one peak. Furthermore, position B is at the peak position of the integrated intensity of C$_2$H($N=$1--0) (Fig.~\ref{fig_lines}), while A and C are situated on the edge of the emission structure. Thus the double-peaked feature at positions A and C should be because two components coincide along the line of sight, and are not due to self absorption. These two components may trace different parts of the PDR. As mentioned above and showed in Fig.~\ref{fig_chspec}, two velocity components are found in CH$_3$OH($2_{0, 2}-1_{0, 1})$A+ line at position B, but we could not fit the spectrum properly, so we exclude this line from the discussion below. For NH$_3$(1, 1), we do not find double peaked features toward these five positions. All the N$_2$H$^+$(1--0) spectra can be fit with one component, except the one from position C. The isolated component of the average spectrum at position C shows a clear double-peaked feature. This feature is shown in the average spectrum but is very difficult to identify in each single spectrum, because the isolated component of the N$_2$H$^+$(1--0) hyperfine transitions we used to determine the number of components to fit is relatively weak, and the double-peaked feature is only shown toward the southeast edge. The redshifted component of the N$_2$H$^+$(1--0) and C$_2$H($N=$1--0) emission at this position shows similar velocities, which indicates that they are associated with each other. 

To estimate the contribution of the non-thermal component in the molecular emission, we assumed the relation $\sigma_{\rm nth}=\sqrt{\sigma_{\rm obs}^2-\sigma_{\rm th}^2-\sigma_{\rm reso}^2}$, where $\sigma_{\rm obs}$ is the measured velocity dispersion, $\sigma_{\rm th}$ the thermal velocity dispersion, and $\sigma_{\rm reso}$ the velocity dispersion introduced by the velocity resolution of our data. For a Gaussian line profile with the FWHM line width, $\Delta v$ is obtained from the fittings, then $\sigma_{\rm obs}=\Delta v/\sqrt{8 \rm{ln}2}$, and $\sigma_{\rm reso}=0.42/\sqrt{8 \rm{ln}2}$. Assuming a Maxwellian velocity distribution, $\sigma_{\rm th}$ can be calculated from $\sqrt{k_{\rm B} T_{\rm k}/(\mu m_{\rm H})}$, where $k_{\rm B}$ is the Boltzmann constant, $\mu$  the molecular weight, $m_{\rm H}$  the mass of the hydrogen atom, and $T_{\rm k}$  the kinetic temperature. We assumed LTE conditions, so that $T_{\rm rot}$ obtained from NH$_3$ observations approximates the kinetic temperature $T_{\rm k}$, and $T_{\rm k}$ is the same for all molecules. The non-thermal to thermal velocity dispersion ratio $\sigma_{\rm nth}/\sigma_{\rm th}$ is listed in Table~\ref{tab_linefit}, where we also list the Mach number $\sqrt{3} \sigma_{\rm nth}/c_{\rm s}$, where $c_{\rm s}$ is the sound speed estimated using a mean molecular weight $\mu=2.3$.

\begin{table*}
\caption{\label{tab_linefit} Fitting results of the spectral lines shown in Fig.~\ref{fig_chspec}.}
\centering
\begin{tabular}{lcccccc}
\hline\hline
\noalign{\smallskip}
Lines& Position & $v_{\rm peak}$  &$\Delta v$ &$\sigma_{\rm nth}$&$\frac{\sigma_{\rm nth}}{\sigma_{\rm th}}$ & $\sqrt{3} \frac{\sigma_{\rm nth}}{c_{\rm s}}$\\
                         &             &(km~s$^{-1}$)&(km~s$^{-1}$) & (km~s$^{-1}$)& & \\
\hline
\noalign{\smallskip}
C$_2$H($N=$1--0)&A$_1$          &--18.8&        0.69   &0.21 &2.4 &1.3\\
                         & A$_2$                        &--17.7&        0.67     &0.20 &2.3 &1.2\\
                        & B                             &--18.7&        1.03    &0.39&4.5 &2.4\\
                        & C$_1$                 &--19.3&        1.27    &0.50&6.7 &3.5\\
                        & C$_2$                 &--17.3&        0.68   &0.21&2.9&1.5\\
\hline                        
\noalign{\smallskip}
NH$_3$(1, 1)    &A                      &--18.2&        1.52     &0.61&5.7 &3.6\\
                        &B                      &--18.8&        1.40     &0.53&5.0 &3.2\\
                        &C                      &--18.4&        1.44     &0.58&6.3 &4.1\\
                        &D                      &--18.7&        0.76     &0.26&3.1 &2.0\\
                        &E                      &--19.1&        0.87     &0.31&3.9 &2.5\\
\hline
\noalign{\smallskip}
N$_2$H$^+$(1--0)&A              &--18.0&  1.57    &0.64&7.7 & 3.8\\
                        &B                      &--18.7&  0.94   & 0.35&4.3&2.1\\                                     
                        &C$_1$          &--18.6&  1.30   & 0.52&7.4&3.6\\
                        &C$_2$          &--17.3&   0.84   &0.30& 4.3&2.1\\
                        &C$_s$\tablefootmark{$\ast$}&--18.1&  1.92    &0.79&11.4&5.6\\        
                        &D                      &--18.7&        0.82  &0.29& 4.6&2.3\\
                        &E                      &--19.1&        1.14    &0.45&7.3 &3.6\\
\hline
\end{tabular}
\tablefoot{
\tablefoottext{$\ast$}{The Results from single Gaussian component fitting.} 
}
\end{table*}

As listed in Table~\ref{tab_linefit},  the ratios between $\sigma_{\rm nth}$/$\sigma_{\rm th}$ for all the lines are all above 2.3. For C$_2$H($N=$1--0) and NH$_3$(1, 1), the non-thermal contribution to the velocity dispersion is more pronounced toward position C than A and B. For N$_2$H$^+$(1--0), the $\sigma_{\rm nth}$/$\sigma_{\rm th}$ ratio toward position C is much higher than B, but comparable to A. Considering the signal-to-noise ratio for N$_2$H$^+$(1--0) line toward A is not very high, we could not rule out the possibility that the spectra contain two velocity components. The spectra of NH$_3$(1, 1) and N$_2$H$^+$(1--0) show that the non-thermal contribution to the line widths toward position E is larger than toward D. For the Mach number $\sqrt{3} \sigma_{\rm nth}$/$c_{\rm s}$ we obtained, all of them are above 1.0. The Mach numbers for the NH$_3$(1, 1) lines toward D and E fit the average Mach number for quiescent starless cores \citep{sanchez-monge2013}. 

Assuming uniform density across the molecular clump, we estimate the virial mass of the starless clump HMSC-E and HMSC-W as $M_{\rm vir}=210R\Delta v^2$~$M_\odot$ \citep{macLaren1988, sanchez-monge2013}, where $R$ is the radius of the main NH$_3$ emission structures which have the column density measurements (top left panel, Fig.~\ref{fig_col}), and $\Delta v$ is the average NH$_3$(1, 1) line width of the starless clump. The virial mass we obtained for HMSC-E is $\sim$50~$M_\odot$ and  $\sim$14~$M_\odot$ for HMSC-W. To derive the LTE gas mass for HMSC-E and HMSC-W, we first estimated the NH$_3$ fractional abundance by comparing the average gas column density and the average NH$_3$ column density of HMSC-E. The average gas column density was determined from the 3~mm continuum map with the same area as HMSC-E shown in the NH$_3$ column density map in Fig.~\ref{fig_col}. To avoid the contamination from the central continuum source, { we excluded the area within the full width 1\% maximum (similar to FWHM but at the 1\% of maximum for a 2D Gaussian function) from the 3~mm continuum peak. We also took some random points in the area we used for the abundance estimation and estimated the local fractional abundance to estimate the error in the average fractional abundance that we derived. The derived NH$_3$ fractional abundance is $\sim6.3\pm4.0\times10^{-9}$. With this abundance, we derived the LTE gas mass of HMSC-E to be $\sim62^{+110}_{-24}$~$M_\odot$. Assuming that HMSC-W and HMSC-E have the same abundance of NH$_3$, we can estimate the LTE gas mass of HMSC-W to be $\sim60^{+100}_{-24}$~$M_\odot$. The error for the LTE mass were estimated by taking only the error from the NH$_3$ fractional abundance into account. The LTE masses we derived for these two clumps are higher than the virial masses, which indicates that these clumps could be gravitationally bound so will probably collapse. This also agrees with the study done by \citet{shimoikura2013} with single-dish observations toward giant molecular clouds showing that clumps without IR clusters are mostly located below the $M_{\rm vir}$=$M_{\rm LTE}$ in the $M_{\rm vir}$ vs $M_{\rm LTE}$ plot. Previous studies show that the NH$_3$ fractional abundance in the cold and dense clouds is a few $10^{-9}$ to $10^{-7}$ \citep[e.g.,][]{tafalla2006,pillai2006,foster2009, friesen2009}. The low NH$_3$ fractional abundance we derive also suggests that the N-bearing molecules might start to be depleted, at least in the northern part of the HMSC-E clump \citep{aikawa2005,flower2006,friesen2009}.}

\section{Discussion}
\label{sec_dis}
\subsection{The nature of the continuum cores}
\label{sec_dis_cont}

\begin{figure*}[htbp]
   \centering
   \includegraphics[width=0.8\linewidth]{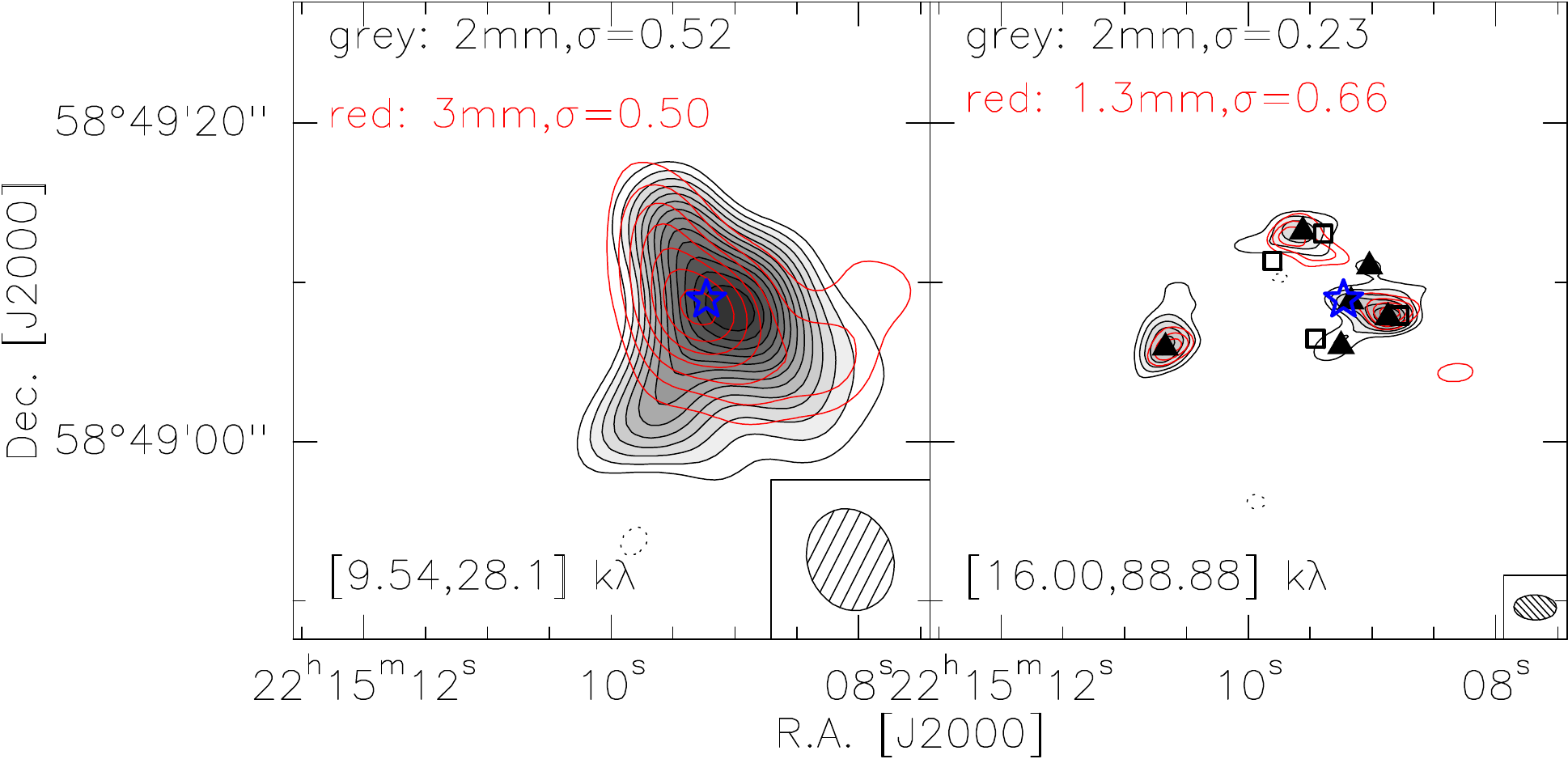}
   \caption{Convolved mm continuum maps. {\it Left:} The 3~mm continuum map (red) overlaid on the 2~mm continuum map (gray scale and black contours), where both maps were produced with the same {\it uv-}range and same beam size.  {\it Right:} The 1.3~mm continuum map (red) overlaid on the 2~mm continuum map (gray scale and black contours), where both maps were produced with the same {\it uv-}range and same beam size. The triangles mark the 2~mm cores, and the squares the 1.3~mm cores detected by \citet{palau2013}. The star indicates the UCH{\sc ii} region VLA1. All the contours in both panels start at 4$\sigma$ and increase in steps of 2$\sigma$. The $\sigma$ value in mJy beam$^{-1}$ and the {\it uv-}range are shown in \textcolor{red}{respect} panel. The dotted contours are the negative features due to the missing flux with the same contour levels as the positive ones in each panel. The synthesized beam is shown in the bottom right corner of each panel.}
     \label{fig_ccont}
\end{figure*}

To compare the continuum emission at different wavelengths, we made 2~mm and 3~mm continuum images with the same range of projected baselines (9.54--28.1~k$\lambda$) and convolved the resulting images to the same resolution of $6\farcs5\times5\farcs4$, P.A.$=18.5\degr$. Similarly we also made 1.3~mm and 2~mm continuum images with the same range of projected baselines (16.00--88.88~k$\lambda$) and convolved the resulting images to the same resolution of 2$\farcs7\times1\farcs6$, P.A.$=88\degr$. The images are shown in Fig.~\ref{fig_ccont}. As we can see in the lefthand panel of Fig.~\ref{fig_ccont}, the 3~mm continuum is consistent with the 2~mm one. Although the 3~mm emission peak is offset by $\sim2\arcsec$ from the 2~mm one, it is smaller than the size of the synthesized beam. The righthand panel in Fig.~\ref{fig_ccont} shows that the 2~mm and 1.3~mm continuum images are consistent with each other. The 1.3~mm continuum emission associated with core MM1 shows extended emission toward VLA1, which is probably filtered out in the original higher spatial resolution map (right panel, Fig.~\ref{fig_mmcont}). With the identification level of 6$\sigma$, \citet{palau2013} detected four continuum sources with higher angular resolution at 1.3~mm with PdBI (open squares in Figs.~\ref{fig_mmcont} and Fig.~\ref{fig_ccont}). One of their cores located in the southwest is associated with MM1. The southeast core from \citet{palau2013} is located close to our continuum core MM4 and is offset by $\sim$1\farcs6 from it. The other two cores in the north could be part of our continuum core MM3. 

Among the dense cores we detected at different wavelengths, MM2 is associated with the UCH{\sc ii} region VLA1, and extended emission at NIR wavelengths. Although the size of VLA1 (0.01~pc) fits a hyper-compact (HC) H{\sc ii}, the SED and the fitting results for the radio emission (Fig.~\ref{fig_sed}) suggest that VLA1 is a UCH{\sc ii} \citep{churchwell2002}, which hosts an intermediate- and high-mass protostellar object. Other continuum sources are located around MM2/VLA1, where MM1 is the strongest source at both 1.3~mm and 2~mm. It is associated with one infrared source and can only be detected in the $K$ band and longer wavelengths, indicating that the source is a deeply embedded intermediate and high-mass protostellar object. MM3a is associated with one IRAC source and shows no emission in the $K $ band, which is considered to be an embedded intermediate- to high-mass protostellar object. All the other continuum sources, MM3b, MM4, MM5 (MM5a and MM5b), and MM6 are not associated with any infrared source, thus all these dense cores could be starless cores. 

Single-dish millimeter measurements for I22134 at 1.2~mm (240~GHz) with IRAM~30~m show a peak intensity of $\sim$ 229~mJy~beam$^{-1}$ with a beam of 11$\arcsec$ \citep{beuther2002a}, and an integrated flux of 2.5~Jy. Assuming a dust emissivity index $\beta=$1.8, that peak intensity corresponds to $\sim$180~mJy~beam$^{-1}$ at 1.3~mm, $\sim$43~mJy~beam$^{-1}$ at 2~mm, and $\sim$5~mJy~beam$^{-1}$ at 3~mm with a beam size of 11$\arcsec$. In our interferometer observations, the flux recovered within a beam of 11$\arcsec$ is 20~mJy at 1.3~mm, meaning that $\sim$90\% of the flux is lost. Thus the masses we derive at 1.3~mm are a lower limit for the current mass of each source. At 2~mm we recover 24~mJy within a beam of 11$\arcsec$ with our PdBI observations. Compared to the single-dish flux ($\sim$43~mJy), we miss $\sim44\%$ of the flux within the IRAM~30~m beam.  However, within the PdBI main beam of 33$\arcsec$ at 2~mm, we only recover $\sim$40~mJy, while the integrated flux reported by \citet{beuther2002a} converts to 2~mm and would be $\sim$400~mJy, indicating that we are filtering out more extend emission. 

At 3~mm, the flux derived from our CARMA observations within the IRAM~30~m beam is 16~mJy, which is much larger than the one we estimated from the 30~m peak flux. Furthermore, the integrated flux reported by \citet{beuther2002a} (2.5~Jy) converts to 3~mm and would be $\sim$56~mJy. The 3~mm flux we recovered within the CARMA main beam of 77$\arcsec$ is $\sim$49~mJy, indicating that we are recovering most of the central and extended emission. The larger peak flux from CARMA might be due to our having underestimated the 3~mm flux with a large $\beta$ and the free-free emission contribution at 3~mm. \citet{williams2004} found a mean $\beta$ of 0.9 toward a sample of high-mass protostellar objects. If we adopt this value and take the free-free emission measured from our VLA observations into account, the peak intensity reported by \citet{beuther2002a} would be $\sim$15~mJy at 3~mm, which is similar to the 3~mm integrated flux within the IRAM~30~m beam derived from the CARMA observations. Furthermore, by inserting the flux measurement from \citet{beuther2002a} into Equation 1 and assuming a dust temperature of 25~K, the derived gas mass is $\sim$259~$M_\odot$.

With a resolution of 0\farcs56$\times$0\farcs49, \citet{palau2013} recovered a flux of $\sim$2.8~mJy at 1.3~mm for MM1, which is about $25\%$ of the flux we recovered from our SMA observations, thus the gas mass we derived is several times higher than what they obtained. Considering the shortest baseline for the observations by \citet{palau2013} is 136~m, any structure larger than 0.9$\arcsec$ would be filtered out \citep{palau2010}, which is the reason they only recovered $25\%$ of the flux we did.

\subsection{Chemistry}
\subsubsection{Deuterium}

Previous observations and models show that deuterium fractionations $D_{\rm frac}$ can be used as good tracers of the evolutionary stages for low-mass dense cores. For earlier starless cores, CO is depleted on dust grains in cold ($T<$20~K) and dense cores ($n\sim10^5$~cm$^{-3}$), which would lead to an enhancement of the abundance of H$_2$D$^+$ and the deuterated molecules formed from it. As soon as the YSO is formed at the core center, it begins to heat its surroundings, the CO evaporates from the dust grains and starts to destroy the deuterated species, thus $D_{\rm frac}$ decreases \citep[e.g.,][N$_2$D$^+$, NH$_2$D]{hatchell1998, roberts2000, crapsi2005, emprechtinger2009}. Single pointing observations toward massive star formation regions indicate that $D_{\rm frac}$(N$_2$H$^+$) is high at the prestellar/cluster stage and then drops as the temperature of the cores increases and the protostellar objects form, similar to the low-mass regions \citep{fontani2011, gerner2015}. On the other hand, $D_{\rm frac}$(NH$_3$) does not show significant differences between different evolutionary stages in high-mass star-forming cores \citep{fontani2015}.

We detected three NH$_2$D cores distributed around the NIR cluster (Fig.~\ref{fig_lines}), all of which are associated with NH$_3$ emission with $T_{\rm rot} \lesssim18$~K (Fig.~\ref{fig_col}). No NH$_2$D emission is detected toward the main NH$_3$ peak of HMSC-E, which could be due to the high temperature and intensive UV radiation from the NIR cluster toward this region ($\gtrsim20$~K, Fig.~\ref{fig_col}). Both main NH$_3$ peaks for HMSC-W share similar temperatures, but no NH$_2$D emission is detected toward the northern peak, which might be because the northern peak is closer to the NIR cluster and suffers from stronger UV radiation from the cluster members (Figs.~\ref{fig_lines} and \ref{fig_col}). From single pointing observations, \citet{fontani2015} obtained a $D_{\rm frac}$(NH$_3$) of 0.04 and 0.057 for HMSC-E and HMSC-W, respectively. With a beam size of $\sim 28\arcsec$, the pointing \citet{fontani2015} performed toward HMSC-E does not cover NH$_2$D-E, and the pointing toward HMSC-W covers about one half of NH$_2$D-W. It is difficult to compare their results with ours, but the $D_{\rm frac}$(NH$_3$) that \citet{fontani2015} obtained agree with the number we derived for these three NH$_2$D cores within an order of magnitude. 

Although the $D_{\rm frac}$(NH$_3$) we derived are much higher than the [D/H] interstellar abundance ($\sim10^{-5}$, \citealt{oliveira2003}), they are still about one order of magnitude lower than the $D_{\rm frac}$(NH$_3$) of pre-protostellar cores derived from other interferometer observations \citep{busquet2010}. Our results show that the NH$_2$D cores share similar temperatures. The one located farthest away from the NIR cluster, NH$_2$D-E, shows the largest $D_{\rm frac}$(NH$_3$), which indicates that the UV radiation forms the NIR cluster and that the UCH{\sc ii} region plays an important role in the NH$_2$D/NH$_3$ chemistry \citep{palau2007a, busquet2010}. 

In contrast to the NH$_2$D cores, the detected N$_2$D$^+$ cores are located very close to the UCH{\sc ii} region and the NIR cluster; i.e., N$_2$D$^+$-C is only at a projected distance of $\sim8~000$~au from the UCH{\sc ii} region VLA1. N$_2$D$^+$-W and N$_2$D$^+$-C are not associated with any N$_2$H$^+$ emission, but these two cores are associated with the central filament identified in NH$_3$. We have shown in Sect.~\ref{sec_kin} that the proximity of these deuterated cores to the UCH{\sc ii} region might just be a projection effect. The $D_{\rm frac}$(N$_2$H$^+$) obtained from single-dish observations for HMSC-E (N$_2$D$^+$-E) and VLA1 (N$_2$D$^+$-C, N$_2$D$^+$-W) are 0.023 and 0.08, respectively \citep{fontani2011}. The average $D_{\rm frac}$(N$_2$H$^+$) we derived are much lower, which for the core N$_2$D$^+$-E is $\sim$0.0035, and the lower limit of $D_{\rm frac}$(N$_2$H$^+$) for the N$_2$D$^+$-W core is $\sim$0.048 and $\sim$0.035 for the core N$_2$D$^+$-C (Table~\ref{tab_dfrac}). 

The disagreement between the $D_{\rm frac}$(N$_2$H$^+$), which might occur because our N$_2$D$^+$ observations filter out much of the extended emission. Section~\ref{sec_dis_cont} shows that 44$\%$ of the continuum flux was filtered out by our PdBI observations. \citet{fontani2011} did find a higher $D_{\rm frac}$(N$_2$H$^+$) toward VLA1 than toward HMSC-E, which agrees with our results. All these features indicate that the relatively high $D_{\rm frac}$ in the UCH{\sc ii} region VLA1 found by \citet{fontani2011} could be just a projection effect of the presence of nearby starless cores that were not destroyed by the expansion of the ionized region and that fall inside the 30~m beam.

In the protocluster I05345, \citet{fontani2008} detected two N$_2$D$^+$ emission structures, which are located $\sim10~000$~au from the nearby young B stars. These N$_2$D$^+$ emission structures extend toward a region where no N$_2$H$^+$ emission is detected above 3$\sigma$, but they are associated with NH$_3$ emission \citep{fontani2012a}. This could be because the NH$_3$ observations in both their and our work are done with the VLA and have much higher sensitivity and {\it uv-}coverage than the N$_2$H$^+$ observations, considering that the rms of our N$_2$H$^+$ observation is $\gtrsim20$ times larger than that of the NH$_3$ observation and $\gtrsim5$ times larger than for the N$_2$D$^+$ observation. An alternative possibility is that N$_2$H$^+$ is mostly destroyed by CO desorbed from grain mantles by heating or outflow from nearby YSOs \citep{busquet2011}. 

\subsubsection{NH$_3$/N$_2$H$^+$ abundance ratios}

While the NH$_3$/N$_2$H$^+$ abundance ratios that we derive for HMSC-E is $\sim15$ and for HMSC-W is $\sim60$, the abundance ratios from single-dish measurements \citep{fontani2011, fontani2015} are $\sim17$ and $\sim28$, respectively. \citet{fontani2015} estimated ammonia column densities with a beam size of $\sim32\arcsec$ and the assumed beam filling factor of 1, while the column densities of N$_2$H$^+$ were derived with a much smaller beam size ($\sim9\arcsec$) and corrected for beam filling factor. We measured the size of HMSC-E and HMSC-W from our NH$_3$(1, 1) integrated intensity map (Fig.~\ref{fig_lines}), and the equivalent diameters are $\sim30\arcsec$ and 18$\arcsec$, respectively. Thus the beam size of the ammonia by \citet{fontani2015} is similar to the size of HMSC-E and much larger than HMSC-W, which might be the reason that the NH$_3$/N$_2$H$^+$ abundance ratio we derive for HMSC-E is similar to the one from \citet{fontani2011, fontani2015}, but the ratio for HMSC-W is much greater than the one from \citet{fontani2011, fontani2015}.

The NH$_3$/N$_2$H$^+$ abundance ratio is reported as a chemical tracer for the evolution of dense cores in low-mass star-forming regions. Low-mass starless cores are reported to have NH$_3$/N$_2$H$^+$ abundance ratios of $\lesssim300$, while when the YSOs form in the center of the molecular cores, the ratio drops to $\sim60-90$ \citep[e.g.,][]{caselli2002b,hotzel2004,friesen2010}. Similar behavior is seen in the high-mass regime; furthermore, the ratio in the starless cores grows as high as $\sim10^4$ in the starless cores \citep[e.g.,][]{palau2007a,busquet2011}.  \citet{fontani2012a} and \citet{busquet2011} propose chemical models to explain the observed enhancement of the NH$_3$/N$_2$H$^+$ abundance ratio in dense starless cores. \citet{busquet2011} suggest that in the starless stage, the CO depletion favors the formation of NH$_3$ over N$_2$H$^+$, and their model shows that the abundance ratio of a collapsing core grows from $\sim0.2$ to $\sim10^4$ within 0.5 million years. HMSC-W shows higher NH$_3$/N$_2$H$^+$ abundance ratio than that in HMSC-E; furthermore, HMSC-W is at virial equilibrium and might be on the verge of gravitational collapsing. All these features suggest that HMSC-W might be more evolved than HMSC-E. While Table~\ref{tab_linefit} shows the line widths toward HMSC-W and HMSC-E are dominated by non-thermal component, HMSC-E shows higher temperature and line width than HMSC-W (Figs.~\ref{fig_col} and \ref{fig_width}), which might be due to it being closer to the UCH{\sc ii} than HMSC-W where the feedback from UCH{\sc ii} is also stronger.

\subsection{The expanding ``bubble''}
\label{sec_bubble}
\begin{figure*}[htbp]
   \centering
   \includegraphics[width=0.4\linewidth]{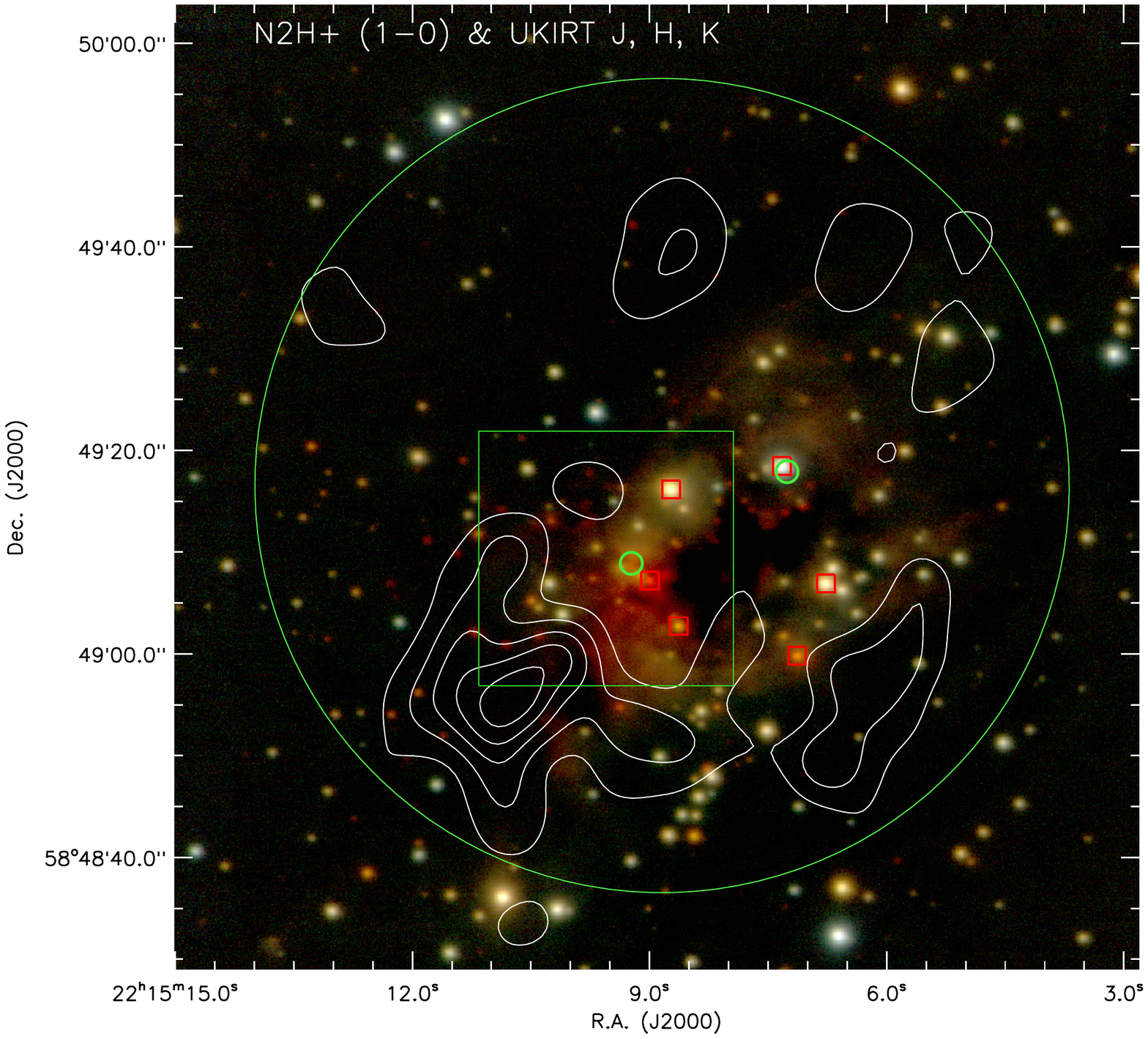}   
   \includegraphics[width=0.4\linewidth]{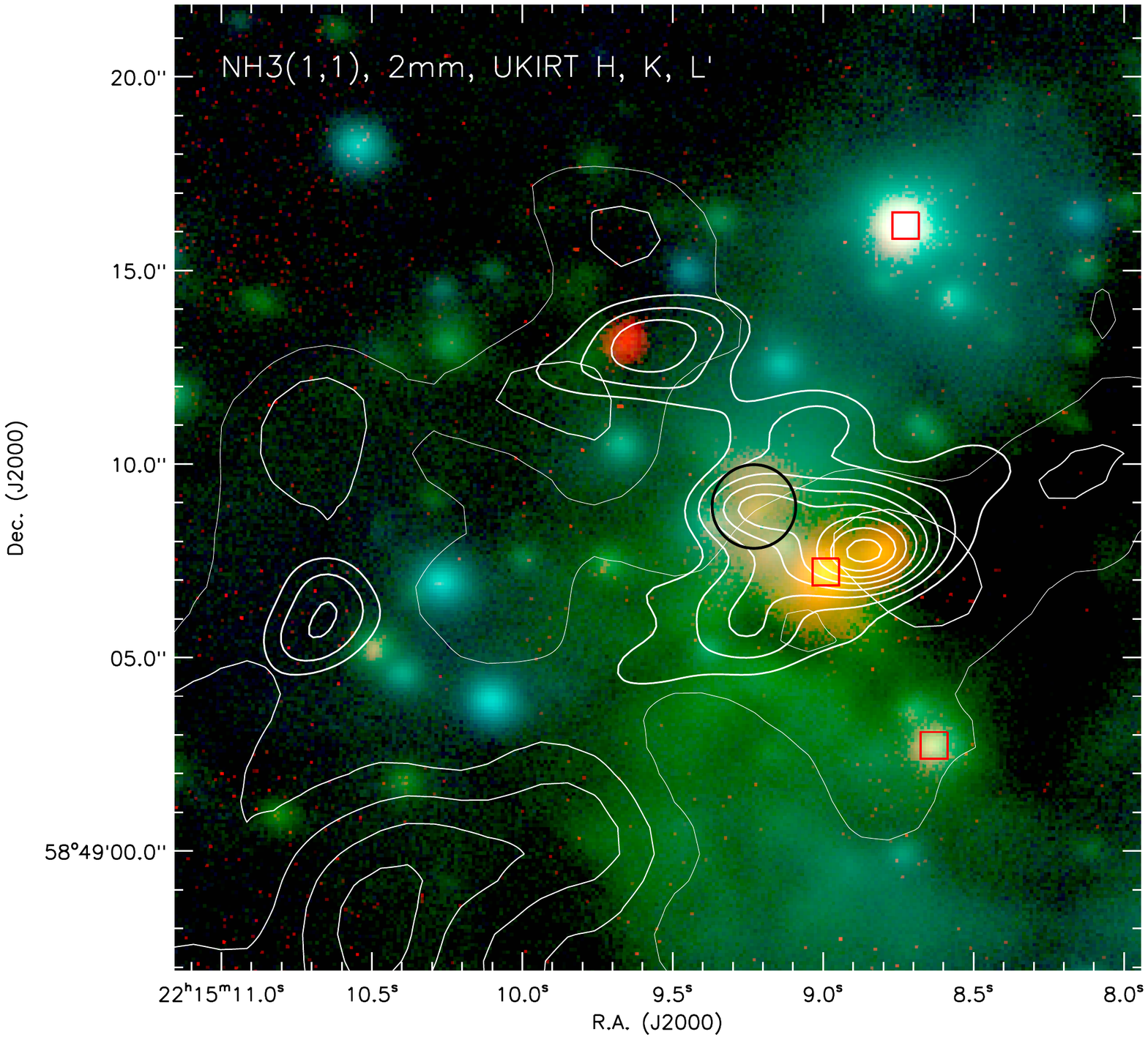}
   \caption{{\it Left:} N$_2$H$^+$(1--0) integrated intensity contours overlaid on $J$ (blue), $H$ (green), and $K$ (red) three-color image obtained from \citet{kumar2003}. Contour levels start from 3$\sigma$ with 6$\sigma$ per level. The big circle indicates the size of the expanding bubble, the small circle in the center the size of the UCH{\sc ii} region VLA1. The red squares label the B stars identified by \citet{kumar2003}. {\it Right:} Zoom into the region shown with the green box in the {\it left} panel, the integrated intensity contours of the NH$_3$(1, 1) main group of hyperfine components (thin contours), 2~mm continuum emission (thick contours) overlaid on $H$ (blue), and $K$ (green), and $L\arcmin$(red) three-color image \citep{kumar2003}. The contour level parameters are the same as in Figs.~\ref{fig_mmcont} and ~\ref{fig_lines}. The red squares indicate the B stars identified by \citet{kumar2003}, and the circle the UCH{\sc ii} region VLA1. }
     \label{fig_nir}
\end{figure*}

As shown in Figs.~\ref{fig_lines} and \ref{fig_nir}, N$_2$H$^+$ and NH$_3$ emission is distributed around the NIR cluster and follows the outline of the cluster well. The velocity structure in the C$_2$H($N=$1--0) emission suggests that the newly formed H{\sc ii} region VLA1 and the NIR cluster are disrupting and dispersing the natal cloud. The elongated large scale CH$_3$OH filament may trace the shock produced by the energetic wind that is pushing the gas away. In the lefthand panel of Fig.~\ref{fig_nir}, we outlined the expanding structure with the big circle and call it the "bubble". 

Zooming into the vicinity of the UCH{\sc ii} VLA1, \citet{palau2013} and \citet{sanchez-monge2013} find that the NH$_3$ emission forms a tilted U-shaped structure around VLA1 (right panel, Fig.~\ref{fig_nir}). Except for MM6 and MM2, the continuum cores detected in this work and \citet{palau2013} are all distributed along the arms of the U-shaped NH$_3$ structure on the edges facing UCH{\sc ii} VLA1. Furthermore, \citet{palau2013} suggest that these continuum cores could be affected by the expanding ionization front of the UCH{\sc ii} VLA1. One arm of the U structure, the central filament, extends across the center of the ring-shaped NIR cluster, which \citet{kumar2003} suggest is an empty cavity. Our results show that the center filament is in the foreground of the cluster, and the ``cavity'' is not empty but might be tracing the absorption from the filament (Fig.~\ref{fig_lines}). We also find that NH$_3$(1, 1) and C$_2$H($N=$1--0) show large line widths toward the U structure (Fig.~\ref{fig_width}), furthermore the PV-diagram pv-1 for C$_2$H($N=$1--0) in Fig.~\ref{fig_nir} shows the C-shaped structure and might be tracing an expanding structure \citep{arce2011}. All these features confirm the results from \citet{palau2013} and the U structure might be tracing expansion of the UCH{\sc ii} region VLA1.

To find out the driving source of the ``bubble'', we first compared the energy of the ``bubble'' and the gravitational binding energy of the molecular cloud following the method outlined by \citet{arce2011}. We taook the expanding velocity $v_{\rm b}$ from the C$_2$H($N=$1--0) PV-diagram to be $\sim1-1.5$~km~s$^{-1}$, the mass derived from single-dish 1~mm continuum measurements 259~$M_\odot$, radius derived from N$_2$H$^+$(1--0) integrated intensity map $\sim40\arcsec$, 0.5~pc. Then the gravitational binding energy is estimated to be $\sim9.5\times10^{45}$~erg. The kinetic energy of the ``bubble'' is $E_{\rm b}\sim2.6-5.7\times10^{45}$~erg, which is not enough to unbind the molecular cloud complex. 

If the ``bubble'' is driven by stellar winds from the B1 ZAMS star, which also drives the UCH{\sc ii} region VLA1, we can then estimate the wind mass loss rate $\dot{m}_{\rm w}$ we need to drive the ``bubble'' \citep{arce2011}: 
$$
\dot{m}_{\rm w}=\frac{P_{\rm b}}{ v_{\rm w} \tau_{\rm w}}
$$

\noindent where $P_{\rm b}$ is the total momentum of the ``bubble'' $\sim259-389$~$M_\odot$~km~s$^{-1}$, $v_{\rm w}$ is the wind velocity that is typically $(2-6)\times10^2$~km~s$^{-1}$ for A and B type stars \citep{lamers1999,lamers1995}, we take 400~km~s$^{-1}$, and $\tau_{\rm w}$ is the the wind timescale $\sim10^6$~yr (the duration that the wind has been active). Then $\dot{m}_{\rm w}$ is estimated to be $\sim(7-10)\times10^{-7}$~$M_\odot$~yr$^{-1}$. For a B1 main sequence star, assuming an effective temperature of 2.4$\times10^4$~K \citep{gray2005}, metallicity Z of 1, we could derive the mass loss rate for the wind from VLA1 to be $\sim$10$^{-6}$~$M_\odot$~yr$^{-1}$ following the routine created by Jorick S. Vink \citep{vink2001, vink2000, vink1999}. Thus VLA1 itself is already powerful enough to drive the ``bubble''. The spectral type of other B stars in the cluster is $\sim$B3 \citep{kumar2003}, and the wind from those B stars would have a mass loss rate of $\sim$10$^{-6}$~$M_\odot$~yr$^{-1}$.  Furthermore, Figure~\ref{fig_nir} shows that VLA2 falls quite close to the geometric center of the big cavity traced by N$_2$H$^+$ emission; therefore, we conclude that with all these B stars (Fig.~\ref{fig_nir}, \citealt{kumar2003}), including UCH{\sc ii} VLA1 and VLA2, the stellar wind from the ring cluster should be strong enough to drive the ``bubble''. Then the wind energy injection rate can be estimated \citep{mckee1989}:

$$
\dot{E}_{\rm w}=\frac{1}{2}(\dot{m}_{\rm w} v_{\rm w})v_{\rm rms}.
$$

The rms velocity of the turbulent motions $v_{\rm rms}$ equals $ \sqrt{3} \sigma_{\rm obs}$. The average line width for our N$_2$H$^+$(1--0) observation is $\sim1.0$~km~s${^{-1}}$, which gives $v_{\rm rms}\sim0.74$~km~s${^{-1}}$. We found that the energy ejection rate needed to drive the ``bubble'' is $\dot{E}_{\rm w}\sim(1.2-1.8)\times10^{32}$~erg~s$^{-1}$, which is similar to what \citet{arce2011} found for the shells in Perseus. 

Another possible driving source(s) of the ``bubble'' are the outflows from the YSOs, as they are considered to have a strong influence on the kinematic properties of the parent clouds \citep[e.g.,][]{nakamura2007}. The total energy of the main outflow (south-north direction) we derived in Sect.~\ref{sec_outflow} is $3.24\times10^{44}$~erg, which is only $\sim$10\% of the gravitational binding energy and is not enough to unbind the molecular cloud complex. From single-dish observations, \citet{beuther2002b} found the total outflow energy to be $3.4\times10^{46}$~erg, which should be enough to drive { the ``bubble''}.

\subsection{General picture}
\label{sec_pic}

\begin{figure*}[htbp]
   \centering
   \includegraphics[width=0.8\linewidth]{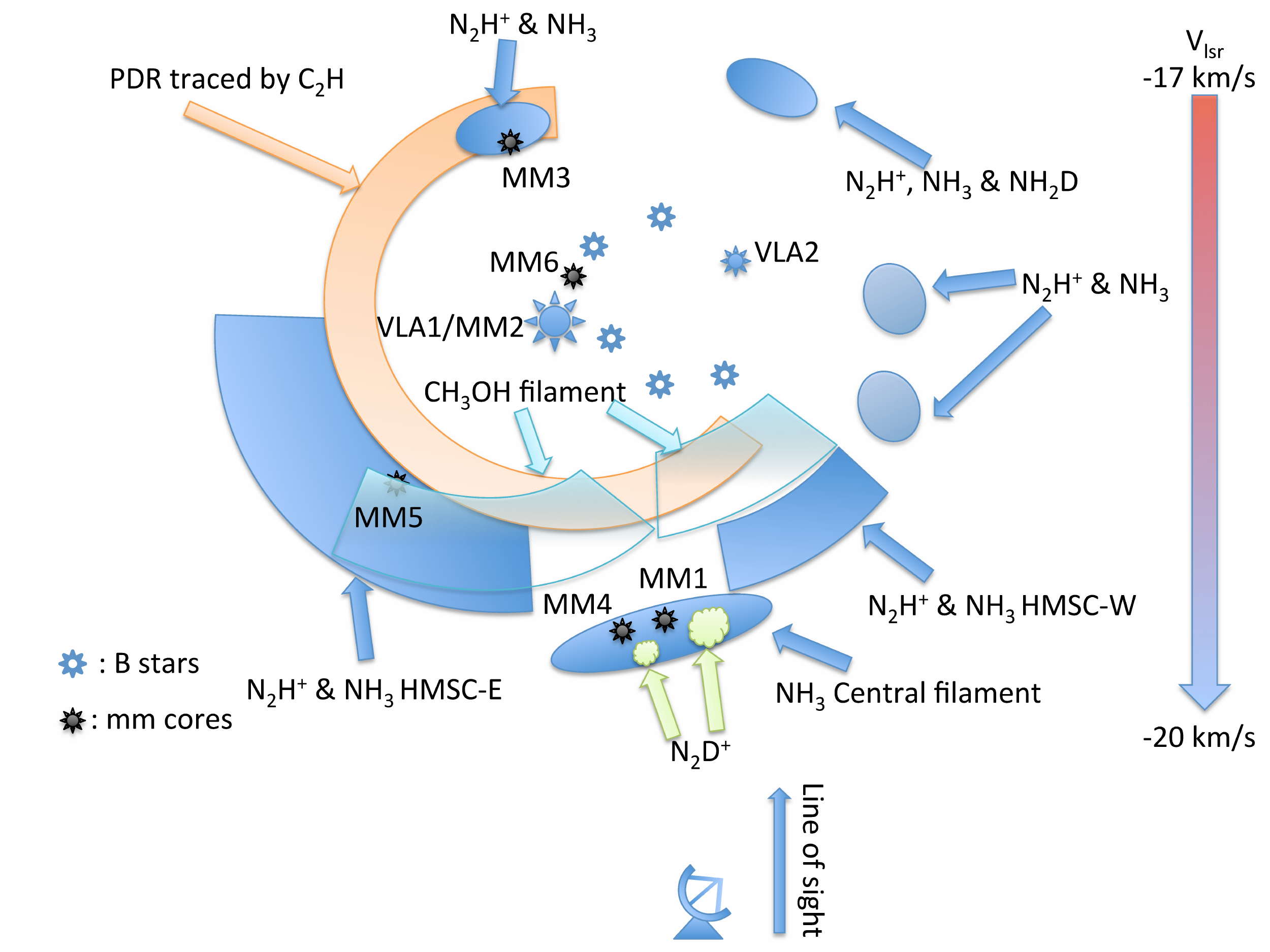}   
   \caption{General picture of I22134 according to the kinetic properties of the molecular line observations, not scaled. The blue ellipses and block indicate the dense clouds traced by different molecules as labeled in the figure. The two green clouds mark the N$_2$D$^+$ cores, the orange arc shows the PDR traced by C$_2$H, the light blue blocks show the possible shocked emission traced by CH$_3$OH, the small stars the B stars in the cluster \citep{kumar2003}, the big filled star the UCH{\sc ii} region VLA1, the small filled star marks VLA2, the black stars the millimeter cores from our observations. The antenna shows our observation point.}
     \label{fig_pic}
\end{figure*}

Combining all the kinematic properties, we obtained from different molecular tracers, we propose a general picture to illustrate the star formation activity in I22134 (see Fig.~\ref{fig_pic}). The UCH{\sc ii} region VLA1, radio source VLA2, and the ring shape B stars \citep{kumar2003} that are part of the NIR cluster sit in the central part of I22134. The PDR traced by C$_2$H($N=$1--0) and other dense clouds traced by different molecules are distributed around the cluster. The detected mm cores are located on the edge the dense clouds facing the UCH{\sc ii}. We propose a possible star formation scenario, where NIR cluster formed first in the same natal cloud, including the YSOs associated with the UCH{\sc ii} regions VLA1 and VLA2. Then the expanding UCH{\sc ii} region VLA1 created the U structure, maybe also triggering the formation of the other mm dense cores, e.g., MM1, MM3a, around VLA1. The NIR cluster and the UCH{\sc ii} region together are disrupting the natal cloud and HMSC-W and HMSC-E around them, may or may not be triggered.

\section{Summary}
\label{sec_sum}
The protocluster associated with I22134 represents an excellent laboratory for studying the influence of massive YSOs on nearby starless cores. We have characterized the different physical and chemical properties of the dense gas by means of VLA, CARMA, PdBI, and SMA observations of the centimeter and millimeter continuum, as well as several molecular tracers (NH$_3$, NH$_2$D, N$_2$H$^+$, N$_2$D$^+$, C$_2$H, CH$_3$OH, CO, SO). We summarize the main results below. 

Multiwavelength (E)VLA observations resolve two radio sources, VLA1 and VLA2. VLA1 is considered to be a UCH{\sc ii} region,  and it shows basically constant flux at centimeter wavelengths. We can fit the SED with a homogeneous optically thin HII region with a size of 0.01 pc, and the ionizing photon flux corresponds to a B1 ZAMS star. Around the UCH{\sc ii} region, one main clump is detected at 3~mm, and six continuum sources are resolved at 2~mm, MM1--MM6. At 1.3~mm, MM1 has a strong counterpart: MM3 is resolved into MM3a and MM3b, and MM5 is resolved into MM5a and MM5b. MM2 is associated with the UCH{\sc ii} region, MM1 are MM3a are associate with SO emission, and NIR point sources are considered to be intermediate-to-high-mass protostellar objects. MM3b, MM4, MM5a, MM5b, and MM6 are not associated with any NIR sources and could be starless cores. We estimated the column density and mass of each source.

On a large scale, N$_2$H$^+$(1--0) and NH$_3$ emission shows an extended structure around the UCHII region and the NIR cluster, and two main structures seen in the N$_2$H$^+$ map are considered to be starless clumps, which we called HMSC-E and HMSC-W. The rotational temperature ($T_{\rm rot}$) map obtained from NH$_3$ observations shows a higher temperature toward HMSC-E ($\sim$25~K) and lower toward HMSC-W ($\sim15$~K). HMSC-E also shows a lower NH$_3$/N$_2$H$^+$ abundance ratio ($\sim$15) than HMSC-W ($\sim60$). By comparing the virial mass and the LTE mass, we found that both HMSC-E and HMSC-W are gravitationally bound and will probably collapse. On a small scale, we detected outflow from $^{12}$CO emission, but the driving source for the outflow is still not clear. For the deuterium molecules, while NH$_2$D cores distributed around the NIR cluster and are all associated with the NH$_3$ emission, which is at a temperature of $\lesssim$18~K, N$_2$D$^+$, cores are associated with the NH$_3$ central filament, close to the UCH{\sc ii} region, with a projected distance of $\sim8000$~au ($d=2.6$~kpc). We found that the NH$_3$ central filament is blueshifted from the rest of the NH$_3$ emission by $\sim1$~km~s$^{-1}$, indicating it is on a slightly different plane from the main clump in space. 

By studying the velocity map of the molecular line emission, we found that the southeastern part of the N$_2$H$^+$(1--0) and NH$_3$(1, 1) emission is redshifted from the main emission peak by $\sim1$~km~s$^{-1}$, which may indicate that the emission structure is expanding. The PV diagram of C$_2$H($N=1-0$) shows a C-shaped velocity structure, which indicates that the emission is tracing an expanding bubble/shell structure. We found that the U-shaped NH$_3$ emission structure might be tracing the expansion of the UCH{\sc ii} region. We found that the molecular cloud around the NIR cluster is also expanding, forming a ``bubble''. We estimated the stellar wind mass loss rate that is needed to produce the expanding structure and suggested that the stellar winds from the NIR cluster are strong enough to drive the ``bubble''. The outflows from the YSOs also have enough energy to dissipate the parent clouds and drive the ``bubble''. Given the kinematic properties of the molecular line emission, we proposed a general picture and a sequential star formation scenario to illustrate the star formation activities in I22134.

\begin{acknowledgements}
The work is supported by the STARFORM Sinergia Project CRSII2\_141880 funded by the Swiss National Science Foundation. Y.W. also acknowledges support by the NSFC 11303097 and 11203081, China. \'A.~S.-M.\ acknowledges support by the collaborative research center project SFB\,956, funded by the Deutsche Forschungsgemeinschaft (DFG). G.B. is supported by the Spanish MICINN grant AYA2011-30228-C03-01 (cofunded with FEDER funds). A.P. acknowledges financial support from a UNAM-DGAPA-PAPIIT IA102815 grant, M\'exico. We thank Dr. Nanda Kumar for providing the NIR UKIRT images. We thank the productive discussion with Prof. Michael Meyer and Dr. Anastasios Fragkos. We acknowledge the  referee Dr. K. Dobashi for improving the manuscript.

\end{acknowledgements}


\bibliographystyle{aa}
\bibliography{wang2015}{}


\end{document}